\newcommand{\PRE}[1]{{#1}} % Use if preprint style
\documentclass[12pt,aps,prd,preprint,tightenlines,superscriptaddress,
amsmath,amssymb,%showpacs,
nofootinbib]{revtex4}
\RequirePackage[colorlinks=true
,urlcolor=blue
,anchorcolor=blue
,citecolor=blue
,filecolor=blue
,linkcolor=blue
,menucolor=blue
,pagecolor=blue
,linktocpage=true
,pdfproducer=medialab
,pdfa=true
]{hyperref}

\usepackage{amsmath,amssymb,amsthm,amsfonts}
\allowdisplaybreaks
\usepackage{slashed} 
\usepackage{graphicx}
\usepackage{subfigure}
\usepackage{dcolumn}% Align table columns on decimal point
\usepackage{hyperref}      % hypertext links %%ARXIV
\usepackage{bm}% bold math
\usepackage{epsfig}
\usepackage{epstopdf}
\usepackage{setspace}
\usepackage[usenames, dvipsnames]{color}
\usepackage{slashed}
\usepackage{comment}
\newcommand{\be}{\begin{equation}\begin{aligned}}
\newcommand{\ee}{\end{aligned}\end{equation}}
\newcommand{\beq}{\begin{equation}}
\newcommand{\eeq}{\end{equation}}
\newcommand{\beqa}{\begin{eqnarray}}
\newcommand{\eeqa}{\end{eqnarray}}

\newcommand{\ifb}{\text{fb}^{-1}}
\newcommand{\iab}{\text{ab}^{-1}}

\newcommand{\kev}{\text{keV}}
\newcommand{\mev}{\text{MeV}}
\newcommand{\gev}{\text{GeV}}
\newcommand{\tev}{\text{TeV}}

\newcommand{\cm}{\text{cm}}
\newcommand{\m}{\text{m}}
\newcommand{\km}{\text{km}}

\newcommand{\yr}{\text{yr}}

\newcommand{\eg}{{\em e.g.}}

\renewcommand{\eqref}[1]{Eq.~(\ref{#1})}

\newcommand{\secref}[1]{Sec.~\ref{sec:#1}}

\newcommand{\Secref}[1]{Section~\ref{sec:#1}}

\newcommand{\figsref}[2]{Figs.~\ref{fig:#1} and \ref{fig:#2}}
\newcommand{\Figref}[1]{Figure~\ref{fig:#1}}

\newcommand{\appref}[1]{Appendix~\ref{sec:#1}}

\newcommand{\lmin}{L_{\text{min}}}
\newcommand{\lmax}{L_{\text{max}}}

\begin{document}

\preprint{UCI-TR-2017-18}

\title{\PRE{\vspace*{1.0in}}
Heavy Neutral Leptons at FASER
\PRE{\vspace*{.4in}}}

\author{Felix Kling}
\email{fkling@uci.edu}
\affiliation{Department of Physics and Astronomy, University of
California, Irvine, CA 92697-4575 USA
\PRE{\vspace*{.1in}}}

\author{Sebastian Trojanowski\PRE{\vspace*{.2in}}}
\email{strojano@uci.edu}
\affiliation{Department of Physics and Astronomy, University of
California, Irvine, CA 92697-4575 USA
\PRE{\vspace*{.1in}}}
\affiliation{National Centre for Nuclear Research,\\Ho{\. z}a 69, 00-681 Warsaw, Poland
\PRE{\vspace*{.4in}}}

%\date{\today}

\begin{abstract}
\PRE{\vspace*{.2in}} 
We study the prospects for discovering heavy neutral leptons at ForwArd Search ExpeRiment, or FASER, the newly proposed detector at the LHC. Previous studies showed that a relatively small detector with $\sim 10~\m$ length and $\lesssim 1~\m^2$ cross sectional area can probe large unconstrained parts of parameter space for dark photons and dark Higgs bosons. In this work we show that FASER will also be sensitive to heavy neutral leptons that have mixing angles with the active neutrinos that are up to an order of magnitude lower than current bounds. In particular, this is true for heavy neutral leptons produced dominantly in $B$-meson decays, in which case FASER's discovery potential is comparable to the proposed SHiP detector. We also illustrate how the search for heavy neutral leptons at FASER will be complementary to ongoing searches in high-$p_T$ experiments at the LHC and can shed light on the nature of dark matter and the process of baryogenesis in the early Universe.
\end{abstract}

%\pacs{}

\maketitle

%%%%%%%%%%%%%%%%%%%%%%%%%%%%%
% Introduction
%%%%%%%%%%%%%%%%%%%%%%%%%%%%%
\section{Introduction}
\label{sec:introduction}

In the past years, the Large Hadron Collider (LHC) has collected an impressive amount of data and placed constraints on a multitude of models for new physics. Most of these searches are targeting high-$p_T$ signatures corresponding to new strongly interacting heavy particles. However, up to now, there has been no discovery of any elementary beyond the standard model (BSM) particle, which has motivated the community to consider a broader range of different physics signatures. In particular, if new particles are light and weakly coupled, they could travel a macroscopic distance before decaying. Searches for such long-lived particles (LLPs) %provides one the most attractive ways to seek for Beyond the Standard Model (BSM) physics~ as it 
are experimentally clean and theoretically well motivated (see, \eg,~\cite{Battaglieri:2017aum}). While current and planned searches for \mev\ to \gev\ range LLPs typically employ beam-dump experiments or meson factories, LLPs would also be produced abundantly in high-energy $pp$ collisions at the LHC. However, they are typically produced with low $p_T$ and therefore will move along the beam pipe and escape detection in the ATLAS~\cite{Aad:2008zzm} and CMS~\cite{Chatrchyan:2008aa} experiments.

In recent papers~\cite{Feng:2017uoz,Feng:2017vli}, we proposed a new experiment searching for LLPs, the ForwArd Search ExpeRiment (FASER) at the LHC. FASER would be placed in the forward direction from the ATLAS or CMS interaction point (IP) and operate concurrently with the ongoing high-$p_T$ searches. In particular, we considered a small size detector ($20~\cm$ to $1~\m$ radius and a $10~\m$ long cylinder) placed in a representative location $400~\m$ away from the IP along the beam axis, possibly in a side tunnel, after the main LHC tunnel enters the arc section and starts to curve. Importantly, the existing LHC infrastructure, including neutral absorbers and magnets deflecting charged particles, would provide natural shielding from various standard model (SM) backgrounds. An additional layer of shielding is provided by the rock and concrete that separate the main tunnel from the side tunnel in which FASER would be placed.

A more detailed study of the LHC infrastructure around the ATLAS IP revealed an attractive location for FASER in the currently unused side tunnel TI18 in the distance about $480~\m$ away from the IP. It is a former service tunnel used by LEP as a connection between the SPS and the main tunnel. In the following, we will perform sensitivity studies for this location.

In this study, we analyze FASER's prospects for discovering heavy neutral leptons (HNLs) as a well-known example of new fermionic particles that can be found in forward searches~\cite{Gorbunov:2007ak, Atre:2009rg, Drewes:2013gca, Drewes:2015iva,Deppisch:2015qwa}. This analysis is complementary to our previous studies. In particular, unlike dark photons, HNLs with mass above $\sim1~\gev$ can be abundantly produced in heavy-meson decays thanks to their mixing with the SM (active) neutrinos. However, in contrast to dark Higgs bosons, HNLs couplings to mesons are not dictated by the quark Yukawa couplings, but rather by the set of unknown Yukawa couplings between HNLs and the active neutrinos, as well as by the CKM mixing parameters. As a result, HNLs can also be efficiently produced in $D$-meson decays, while less abundantly produced $B$ mesons play a dominant role only when other channels are kinematically forbidden.

Alternatively, one can also consider a detector in some near location inside the straight intersection of the main LHC tunnel between the two beam pipes after they split. The location that is as close as possible to the IP is right behind the neutral particle absorber TAN (Target Absorber Neutral) which we chose as a representative case in~\cite{Feng:2017uoz,Feng:2017vli}. The expected signal of new physics in this case can be enhanced with respect to the far location in models in which the lifetime of the LLPs is too small to reach the far location. This is, \eg, true for a dark photon search if its kinetic mixing with the SM photon is not suppressed too much as discussed in~\cite{Feng:2017uoz}. On the other hand, for a sufficiently large lifetime this advantage is missing, while the far location suffers much less from the expected background and can additionally benefit from allowing more space for a larger detector as we illustrated in the case of dark Higgs boson~\cite{Feng:2017vli}. 

As we will see below, the lifetime of HNLs is typically large and they can easily overshoot the detector at the near location once additionally boosted. In the following, we will therefore focus exclusively on the detector placed at far location in the TI18 tunnel.%, while the analysis for the near location can be performed in the future if deemed necessary.

The paper is organized as follows. In \secref{theory}, we review the basic properties of HNLs. We discuss possible production channels of sterile neutrinos at the LHC as well as their decays that give rise to a signal in FASER in \secref{prodanddec}. The main results of this study with the sensitivity reach of FASER in the searches for HNLs are presented in \secref{reach}. \Secref{beyond} is devoted to discussion of selected scenarios going beyond the minimal seesaw mechanism that can be probed by FASER. We conclude in \secref{conclusions}. A more detailed discussion of fragmentation functions for $D$ and $B$ mesons can be found in \appref{fragmentation}.

%%%%%%%%%%%%%%%%%%%%%%%%%%%%%
% Theory
%%%%%%%%%%%%%%%%%%%%%%%%%%%%%

\section{Properties of heavy neutral leptons\label{sec:theory}}

Undoubtedly, the most important reason to add new heavy neutral leptons to the SM is to explain the neutrino oscillations~\cite{Fukuda:1998mi,Ahmad:2002jz} since they provide an elegant way to generate nonzero neutrino masses via the seesaw mechanism~\cite{Minkowski:1977sc, GellMann:1980vs, Mohapatra:1979ia, Yanagida:1980xy, Schechter:1980gr}. In type-I seesaw models one extends the SM by adding $\mathcal{N}$ neutral right-handed fermions (identified with HNLs) that couple to the SM neutrinos similarly to the coupling between left- and right-handed components of the charged leptons. Since additional right-handed fermions are SM singlets, also Majorana mass terms are allowed, leading to the following Lagrangian,
\begin{equation}
\mathcal{L} = \mathcal{L}_{\rm SM} + i\,\bar{\widetilde{N}}_I\slashed{\partial}\widetilde{N}_I-F_{\alpha I}\bar{L}_\alpha\,\widetilde{N}_I\,\tilde{\Phi}-\frac{1}{2}\bar{\widetilde{N}}^c_I\,M_{I}\,\widetilde{N}_I + {\rm h.c.},
\label{eq:lagrangian}
\end{equation}
where $\widetilde{N}_{I=1,\ldots,\mathcal{N}}$ are the right-handed HNLs and  we will assume $\mathcal{N}=3$, $\tilde{\Phi}_i = \epsilon_{ij}\Phi^{\ast}_j$, $\Phi$, and $L_{\alpha=e,\mu,\tau}$ are Higgs and lepton doublets, $F_{\alpha I}$ are the Yukawa couplings between HNLs and the SM leptons; and $M = \text{diag}(M_1,M_2,M_3)$ is the Majorana mass matrix for HNLs. 

After electroweak symmetry breaking (EWSB), the Higgs field gets a nonzero vacuum expectation value, $v=174~\gev$, and terms in \eqref{eq:lagrangian} proportional to Yukawa couplings generate mixing between HNLs and active neutrinos. The full $6\times 6$ neutrino mass matrix reads
\begin{equation}
M_\nu=
\begin{bmatrix}
0 & M_D\\
M^T_D & M
\end{bmatrix},
\label{eq:matrix}
\end{equation}
where $(M_D)_{\alpha I} = v\,F_{\alpha I}$. 
One then obtains nonzero masses for the SM neutrinos after diagonalization. 

Similarly, the HNL mass eigenstates, $N_I$ (denoted without a tilde), get small contributions from active states with mixing angles given by \begin{equation}
U_{\alpha I}^2 \simeq \frac{|F_{\alpha I}|^2\,v^2}{M_I^2},
\label{eq:mixing}
\end{equation}
where typically $U_{\alpha I}\ll 1$. HNLs inherit their interactions with the SM particles from the active neutrinos with couplings suppressed by the mixing angles. In particular, the
charged current interaction for the HNL is given by $-(g/\sqrt{2})\, W_\mu \bar{\ell}_{L,\alpha} \gamma^\mu U_{\alpha I} N_I + h.c.$. On the other hand, their masses, $m_{N_I}$, are mostly governed by Majorana masses $M_I$ with only small corrections from mixing. Hence, they are often referred to as \textsl{sterile neutrinos} or simply \textsl{heavy neutrinos}. 

If sterile neutrinos are the only particles added to the SM to explain the active neutrino masses, then the presence of at least two heavy neutrinos is required to generate two measured mass differences between the active neutrinos. At least three heavy neutrinos are required if also the lightest active neutrino winds up being massive. In this case, the HNL sector has a three-generation structure similarly to the SM.

However, not all these sterile neutrinos are necessarily within the reach of FASER. Hence, when discussing the sensitivity reach, we will focus for simplicity on a single sterile neutrino, $N$, with mass $m_N$ and mixing angles with the active neutrinos denoted by $U_{eN}$, $U_{\mu N}$ and $U_{\tau N}$ for the electron, muon and tau neutrino, respectively. In addition, for the purpose of presenting results, we will assume that $N$ mixes exclusively with only one of the active neutrinos. We will therefore study the reach of FASER for three scenarios in which the mixing is dominated by only one nonzero value of $U_{\ell N}$ in each case, where $\ell=e,\mu,\tau$. On the other hand, the reach for scenarios in which more than one HNL has mass and couplings in the region of the parameter space covered by FASER, or when the mixing between $N$ and active neutrinos has a more complicated pattern, can be deduced from our results based on the plots presenting the number of events in each of the three cases considered by us. The effective parameters that we vary are then the following:
\begin{equation}
m_N,\hspace{0.5cm}U_{eN},\hspace{0.5cm}U_{\mu N},\hspace{0.5cm}U_{\tau N},\hspace{0.5cm}\textrm{where only one $U_{\ell N}\neq 0$ at a time}.
\end{equation}

The allowed region for the mass of sterile neutrinos spans over more than fifteen orders of magnitude (see, \eg, \cite{Drewes:2015iva} for a recent review). Theoretically, the most appealing, but experimentally extremely challenging, are seesaw models with $M_I$ close to the scale of Grand Unified Theories (GUTs) (see, \eg, \cite{Babu:1992ia}) that allow one to naturally fit the experimental neutrino data with Yukawa couplings $F_{\alpha I}\sim 1$. For much lower $M_I$, one naively expects $F_{\alpha I}\ll 1$, which would keep sterile neutrinos beyond the reach of low-energy experiments. However, for more than one generation of HNLs, accidental or symmetry driven cancellations between contributions to the active neutrino mass matrix coming from different $N_I$s allow larger values of Yukawa couplings to be considered in \eqref{eq:mixing}~\cite{Wyler:1982dd, Mohapatra:1986bd, Branco:1988ex, Barr:2003nn, Malinsky:2005bi, Shaposhnikov:2006nn}. As a result, one can effectively study HNLs with the $N_I$ masses and mixing angles as independent parameters, while still preserving the primary motivation for such models that come from explaining the active neutrino data. 

In order to explain the masses of active neutrinos via the seesaw mechanism, the mixing parameters $U_{\ell N}$ need to be sufficiently large. If light enough, sterile neutrinos could then thermalize in the early Universe and distort successful predictions of the Big Bang Nucleosynthesis (BBN) by contributing to the number of relativistic degrees of freedom~\cite{Hernandez:2013lza}. They could also cause an additional entropy production from their late-time decays. This can be circumvented, if HNLs decay before BBN. Combining this and current experimental bounds, one obtains an effective lower limit on the sterile neutrino mass $m_N\gtrsim 140~\mev$ in the range of the mixing angles that can be covered by searches for LLPs. More detailed studies~\cite{Dolgov:2000jw, Dolgov:2000pj, Ruchayskiy:2012si} show that in some cases $m_N$ can be lowered down to $\mathcal{O}(10~\mev)$.

On the other hand, BBN bounds can be evaded for an even much lighter sterile neutrino, provided that its mixing angles with the active neutrinos are suppressed. This gives rise to a sterile neutrino dark matter (DM) candidate with mass at the $\sim \kev$ scale that was, \eg, proposed as a possible explanation to a $3.5~\kev$ excess~\cite{Bulbul:2014sua, Boyarsky:2014jta} in the XMM-Newton data. One can both accommodate the observed active neutrino data, as well as keep $N_1$ as a DM candidate, if heavier HNLs, $N_{2,3}$, have larger mixing angles and satisfy the aforementioned lower mass limit from BBN. In particular, this is true for the renowned $\nu$MSM model~\cite{Shaposhnikov:2006nn}. In such a case, while an effectively stable $N_1$ would remain outside the reach of FASER, the displaced decays of heavier sterile neutrinos could give rise to an observable signal. For other models motivating the search for \gev-scale sterile neutrinos see, \eg,~\cite{Appelquist:2002me,Appelquist:2003uu}.

One additional motivation that lies behind low-scale seesaw models is that, similarly to high-scale models, they can also explain the baryon asymmetry of the Universe (BAU) via thermal leptogenesis~\cite{Asaka:2005pn}. The lepton asymmetry in this case is generated from CP violation in out-of-equilibrium production and subsequent evolution of sterile neutrinos with oscillation effects taken into account. This requires low values of Yukawa couplings $F_{\alpha I}$s. As a result, the mixing angles are typically of order $U_{\ell N}\sim 10^{-5}-10^{-3}$ (see, \eg,~\cite{Canetti:2012kh} for an extensive discussion) and lie beyond the reach of current experiments. A new generation of dedicated LLP searches, including FASER, is therefore needed to study this scenario.

%%%%%%%%%%%%%%%%%%%%%%%%%%%%%
% Production and Decay
%%%%%%%%%%%%%%%%%%%%%%%%%%%%%

%%%%%%%%%%%%%%%%%%%%%%%%%%%%%
\section{Heavy neutral production and decays\label{sec:prodanddec}}

Sterile neutrinos can be produced at the LHC in decays of mesons and tau leptons~\cite{ Gorbunov:2007ak}, heavy baryon decays~\cite{Ramazanov:2008ph}, as well as in production from on- or off-shell gauge or Higgs bosons (see, \eg,~\cite{BhupalDev:2012zg,Das:2016hof} for a recent discussion). Given our focus on very forward-going light HNLs, the main production mechanism is via meson decays. The dominant contribution comes from leptonic and semileptonic decays of pseudoscalar $D$ and $B$ mesons, while decays of vector mesons are subdominant~\cite{Gorbunov:2007ak}. In addition, tau decays to HNLs are taken into account for a scenario in which a sterile neutrino mixes exclusively with the tau neutrino. Kaon decays into sterile neutrinos are kinematically allowed only in the mass range at which strong constraints exist from previous beam-dump experiments. Hence, they will not play important role in our analysis. We do not consider charged pion decays into HNLs since they only become possible for $m_N\lesssim m_\pi$, which is disfavored by strong cosmological bounds, as discussed above. Importantly, although two-body leptonic decays of mesons are chiral suppressed in the SM, this suppression is partially overcome if the sterile neutrino mass is heavy enough since its mass, $m_N$, replaces the mass of a charged lepton in the chiral suppression factor~\cite{Shrock:1980vy}.

Contributions from baryon decays are subdominant in the case of charmed baryons and typically also small for bottom baryons, although in some regions of parameter space, they can be as large as $15\%$ of the total production rate~\cite{Ramazanov:2008ph}. We will neglect them hereafter given larger uncertainties that are expected in our signal rates.

%%%%%%%%%%%%%%%%%%%%%%%%%%%%%
\subsection{HNL production in meson decays}

To determine the momentum and angular distribution of sterile neutrinos produced in the forward region at the LHC, we first simulate meson production in high-energy $pp$ collisions at large pseudorapidities.
In particular, we use Monte-Carlo event generator EPOS-LHC~\cite{Pierog:2013ria}, as implemented in the CRMC simulation package~\cite{CRMC}, to simulate the kaon distributions. Distributions of heavier mesons are simulated using the FONLL tool~\cite{Cacciari:1998it, Cacciari:2012ny, Cacciari:2015fta} in which differential cross sections for $c$- and $b$-quark production in high-energy $pp$ collisions are calculated.  We use the CTEQ $6.6$ pdfs with $m_b = 4.75~\gev$. Subsequent hadronization is performed with nonperturbative BCFY fragmentation functions~\cite{Braaten:1994bz} for $D$ mesons and Kartvelishvili et al. distribution with fragmentation parameter $\alpha= 24.2$ for $B$ mesons~\cite{Kartvelishvili:1977pi, Cacciari:2005uk}. A more detailed discussion of fragmentation is given in \appref{fragmentation}.

\begin{figure}[tb]
\centering
\includegraphics[width=0.32\textwidth]{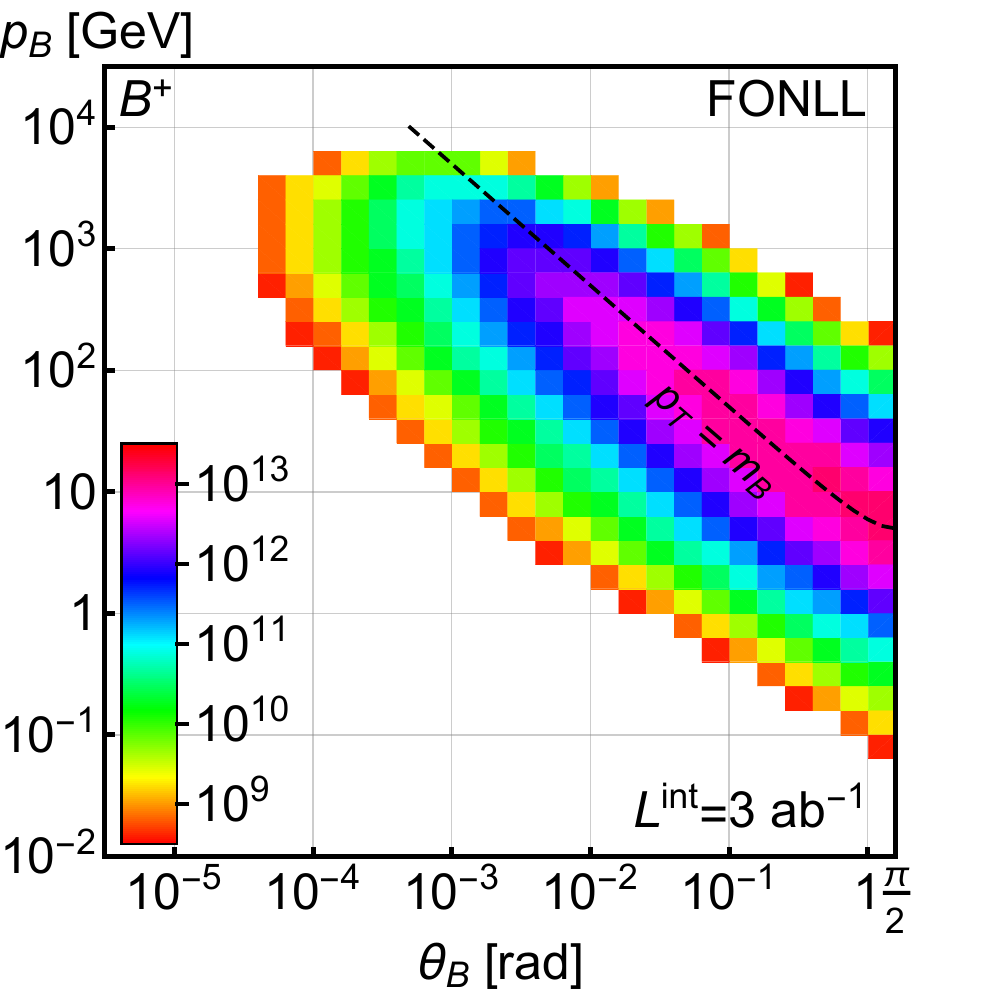}
\includegraphics[width=0.32\textwidth]{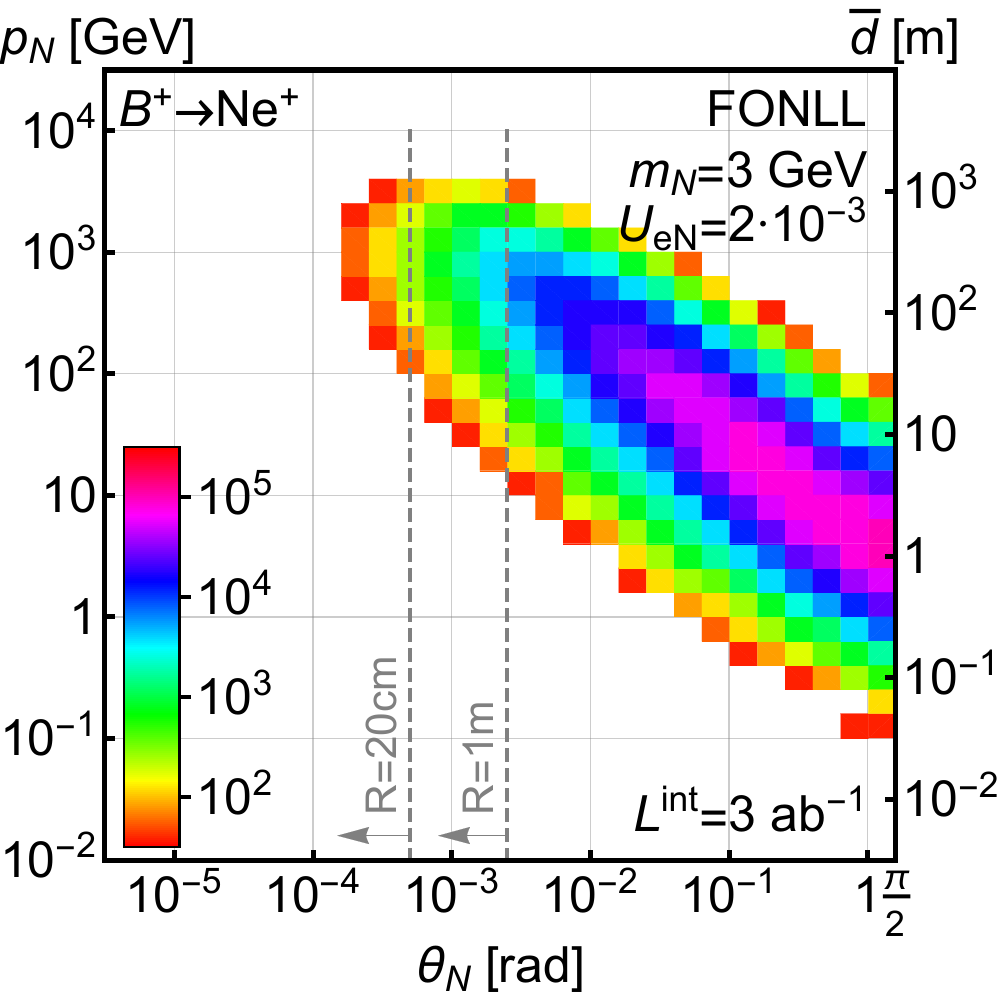}
\includegraphics[width=0.32\textwidth]{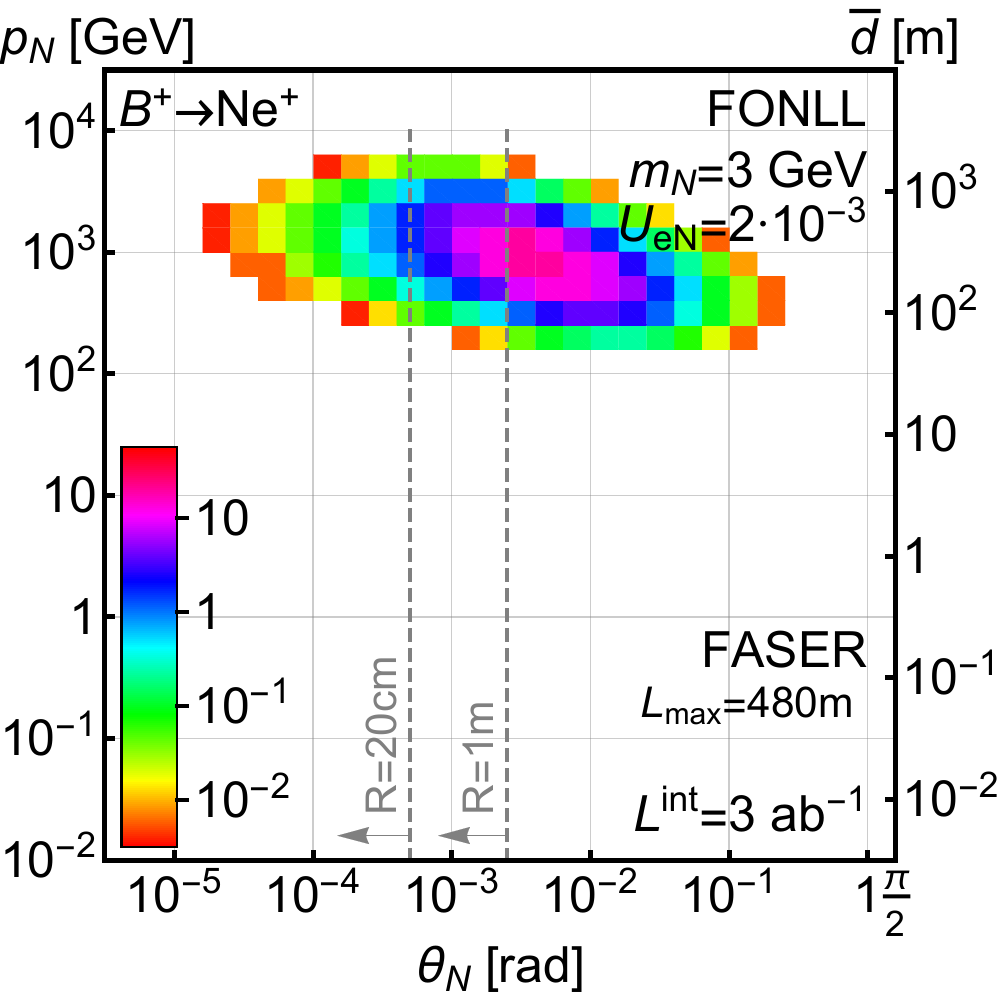}\\
\includegraphics[width=0.32\textwidth]{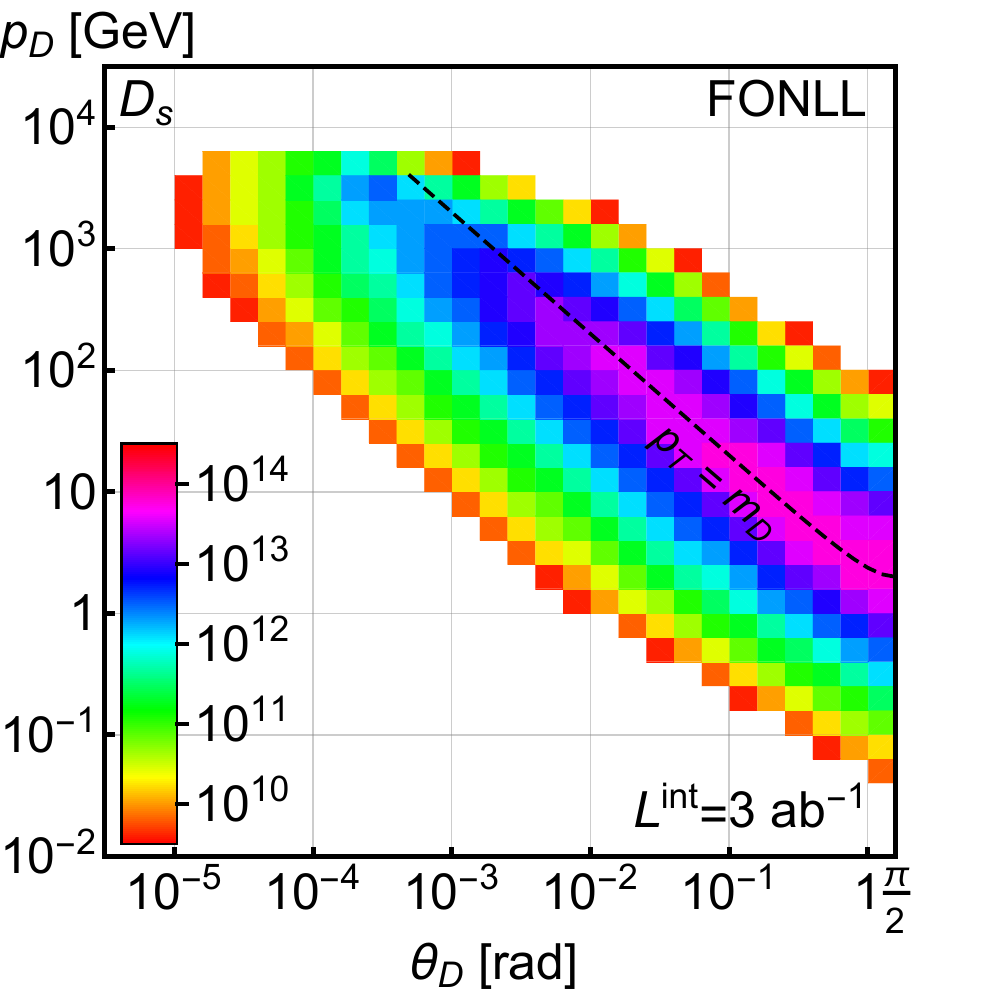}
\includegraphics[width=0.32\textwidth]{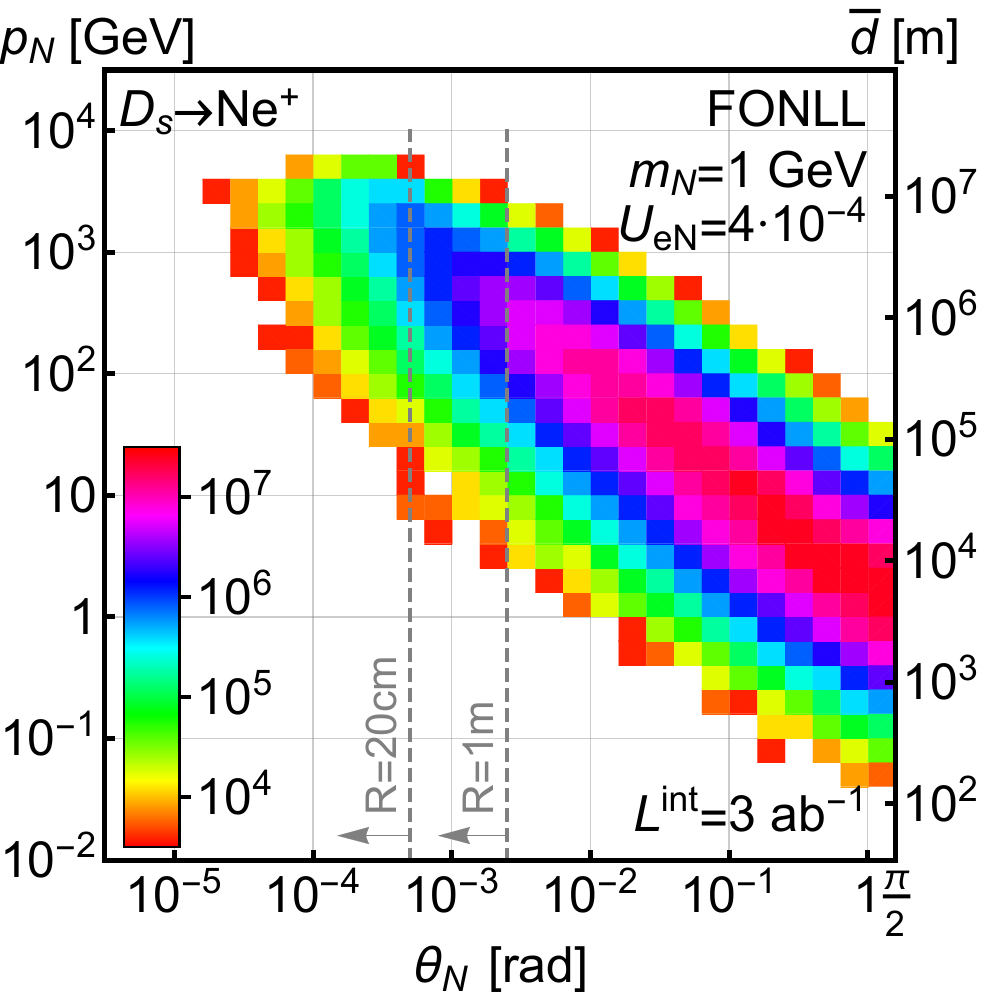}
\includegraphics[width=0.32\textwidth]{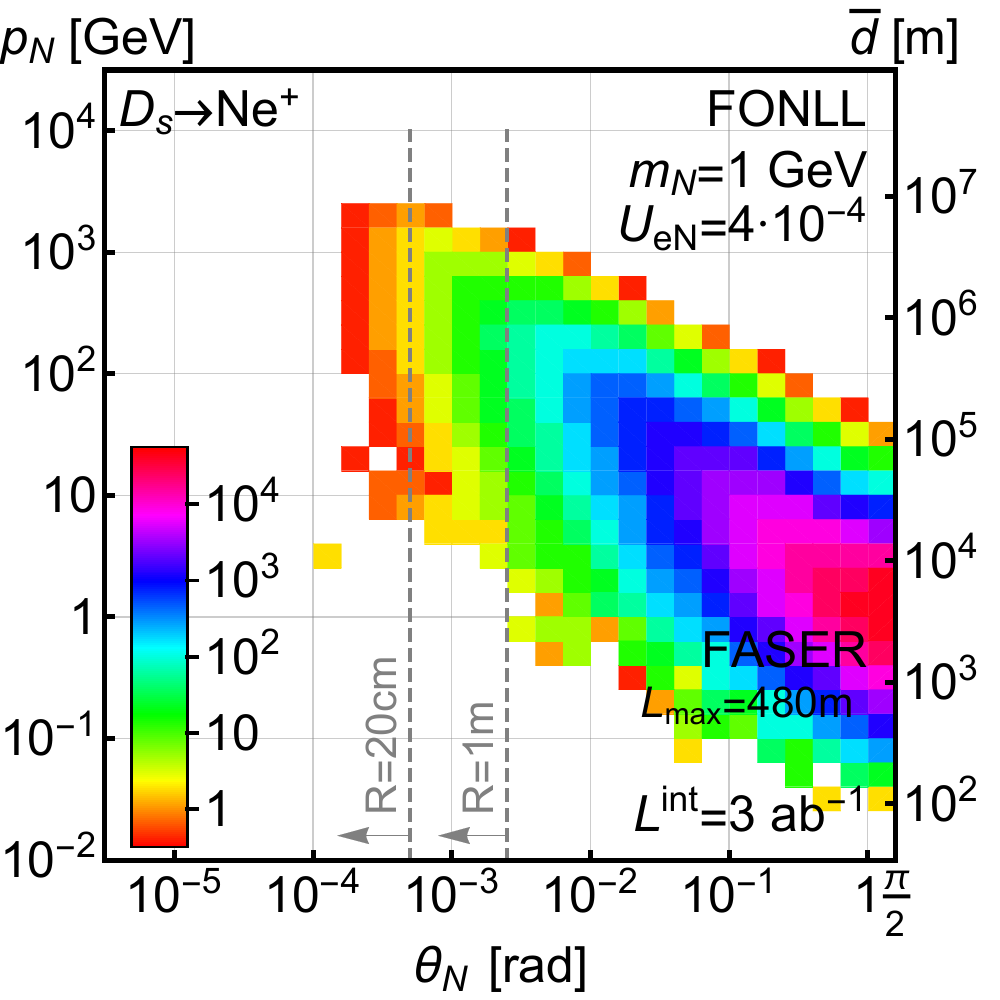}
\caption{Distribution of particles produced at the 13 TeV LHC in the $(\theta, p)$ plane, where $\theta$ and $p$ are the particle's angle with respect to the beam axis and momentum, respectively.  The panels show the number of particles produced in one hemisphere ($0 < \cos \theta \le 1$) for an integrated luminosity of $3~\iab$.  The bin thickness is $1/5$ of a decade along each axis.  The top row shows the distributions of $B^\pm$ mesons (left), heavy neutral leptons produced in $B^\pm \to e^\pm N$ decays (center), and heavy neutral leptons produced in $B^\pm$ decays that themselves decay after traveling a distance in the range $(\lmin, \lmax) = (470~\m, 480~\m)$ (right) for model parameters $(m_\phi, \theta) = (3~\gev, 2\cdot10^{-3})$. The bottom row shows the analogous distributions for $D_s^\pm$ mesons and $(m_N, U_{eN}) = (1~\gev,4\cdot10^{-4})$. The black dashed line corresponds to $p_T = p \sin \theta =  m_B$ in the top left plot and $m_D$ in the bottom left plot. The gray vertical dashed lines in the center and right plots correspond to the angular size of the FASER detector for two choices of its radius considered in the main text: $R=20~\cm$ and $R=1~\m$.}
\label{fig:PvsT}
\end{figure}

In the top left and bottom left panels of \Figref{PvsT}, we show the distribution of $B^\pm$ and $D_s^\pm$ mesons in the $(\theta,p)$ plane, where $\theta$ and $p$ are the meson's angle with respect to the beam axis and their momentum, respectively. The kinematic distributions are clustered around $p_T\sim m_B$ and $m_{D_s}$, respectively. The total number of mesons produced in one hemisphere at $13~\tev$ LHC with an integrated luminosity of $3~\iab$ is $1.4\times 10^{15}$ for $B$ mesons~
\cite{Cacciari:2012ny, Aaij:2016avz} and $1.9\times 10^{16}$ for $D$ mesons~\cite{Cacciari:2012ny, Aaij:2015bpa}.

\begin{figure}[t]
\centering
\includegraphics[width=0.32\textwidth]{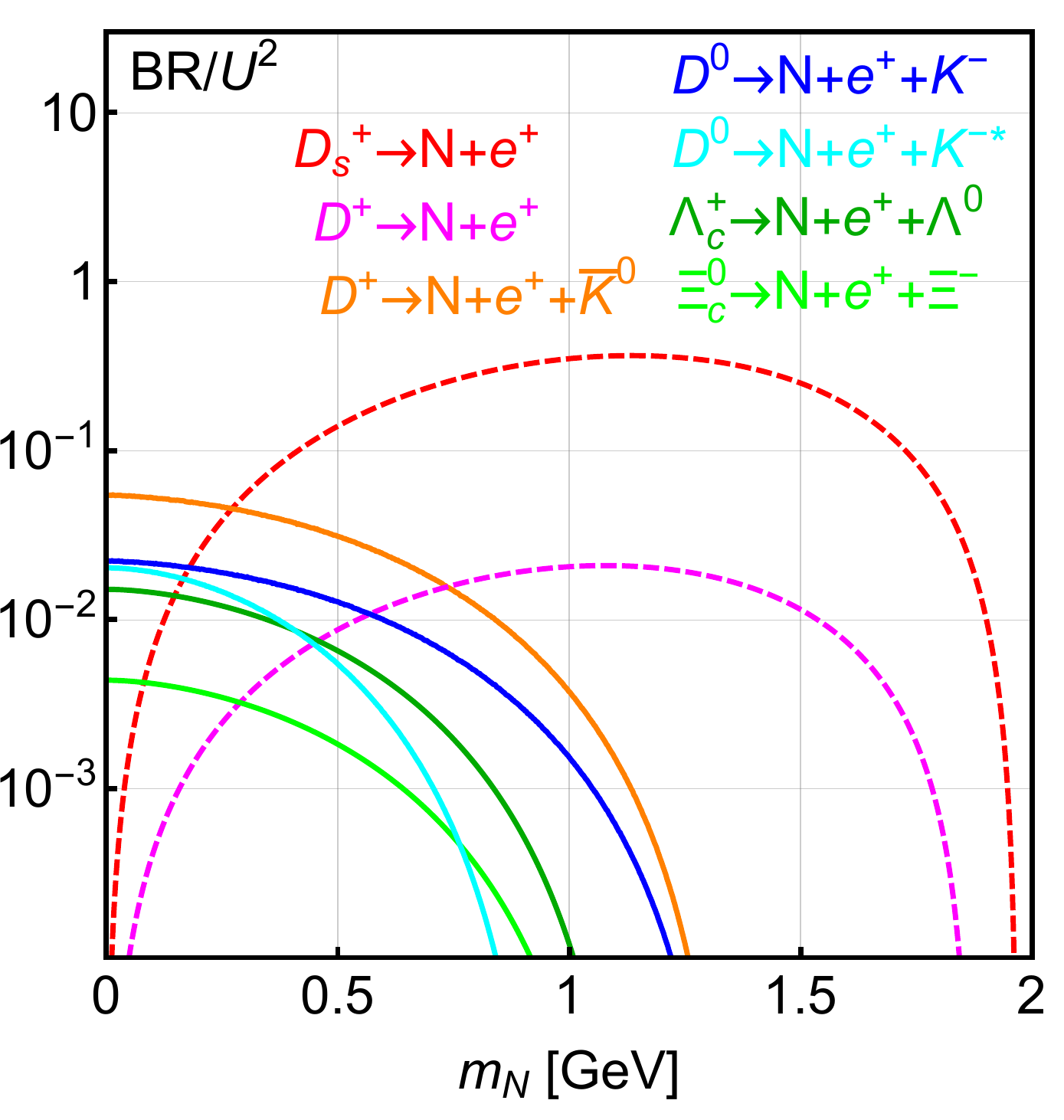}
\includegraphics[width=0.32\textwidth]{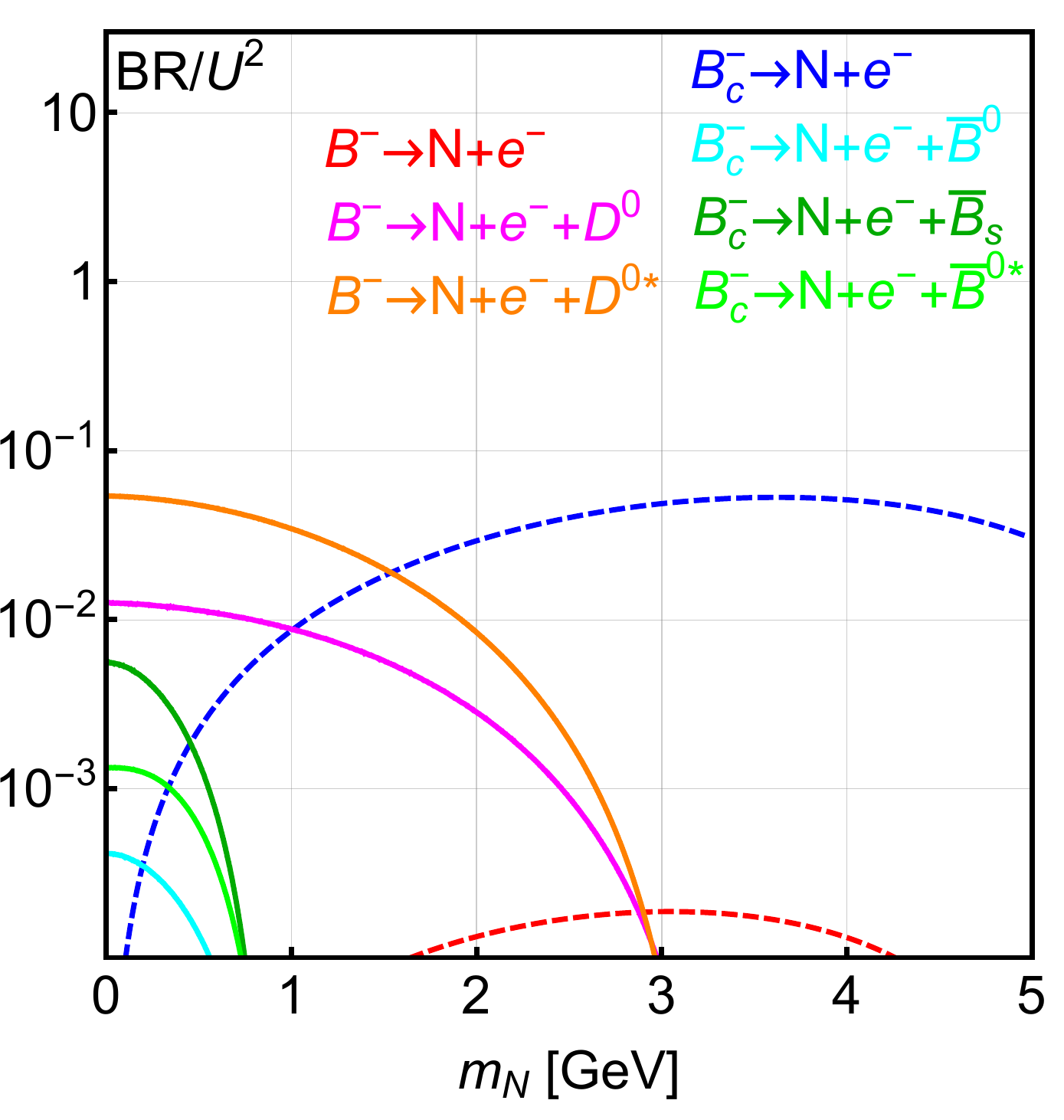}
\includegraphics[width=0.32\textwidth]{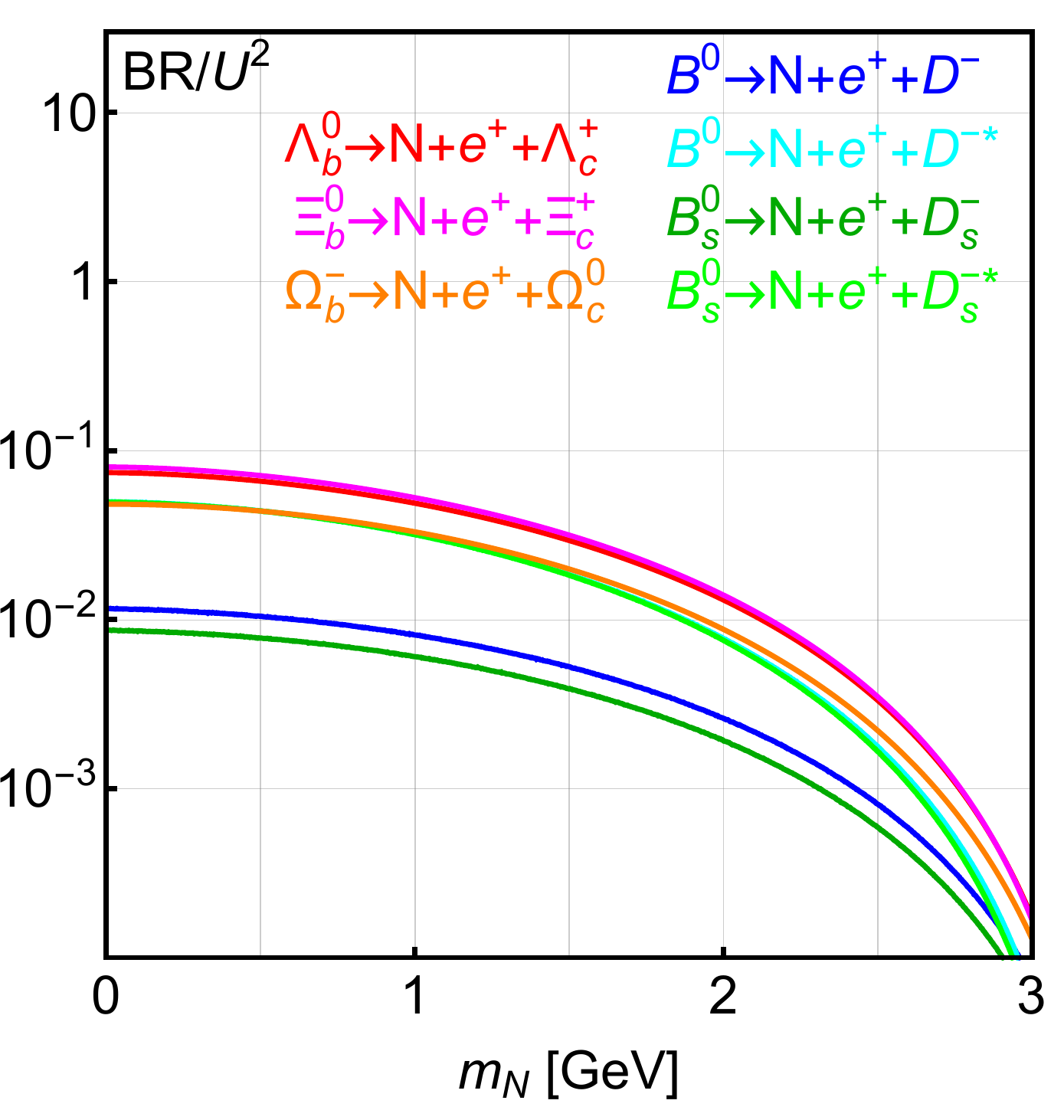}\\
\includegraphics[width=0.32\textwidth]{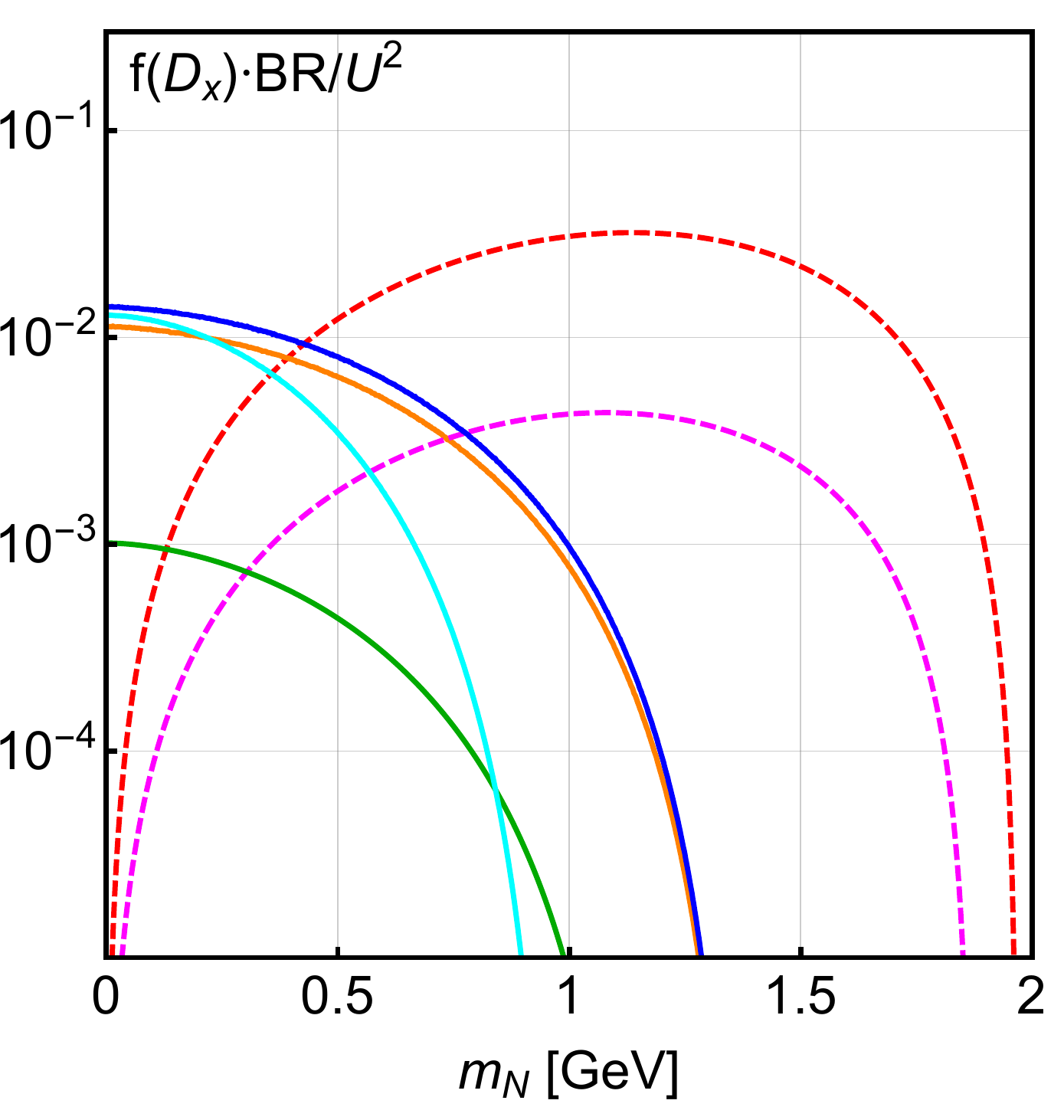}
\includegraphics[width=0.32\textwidth]{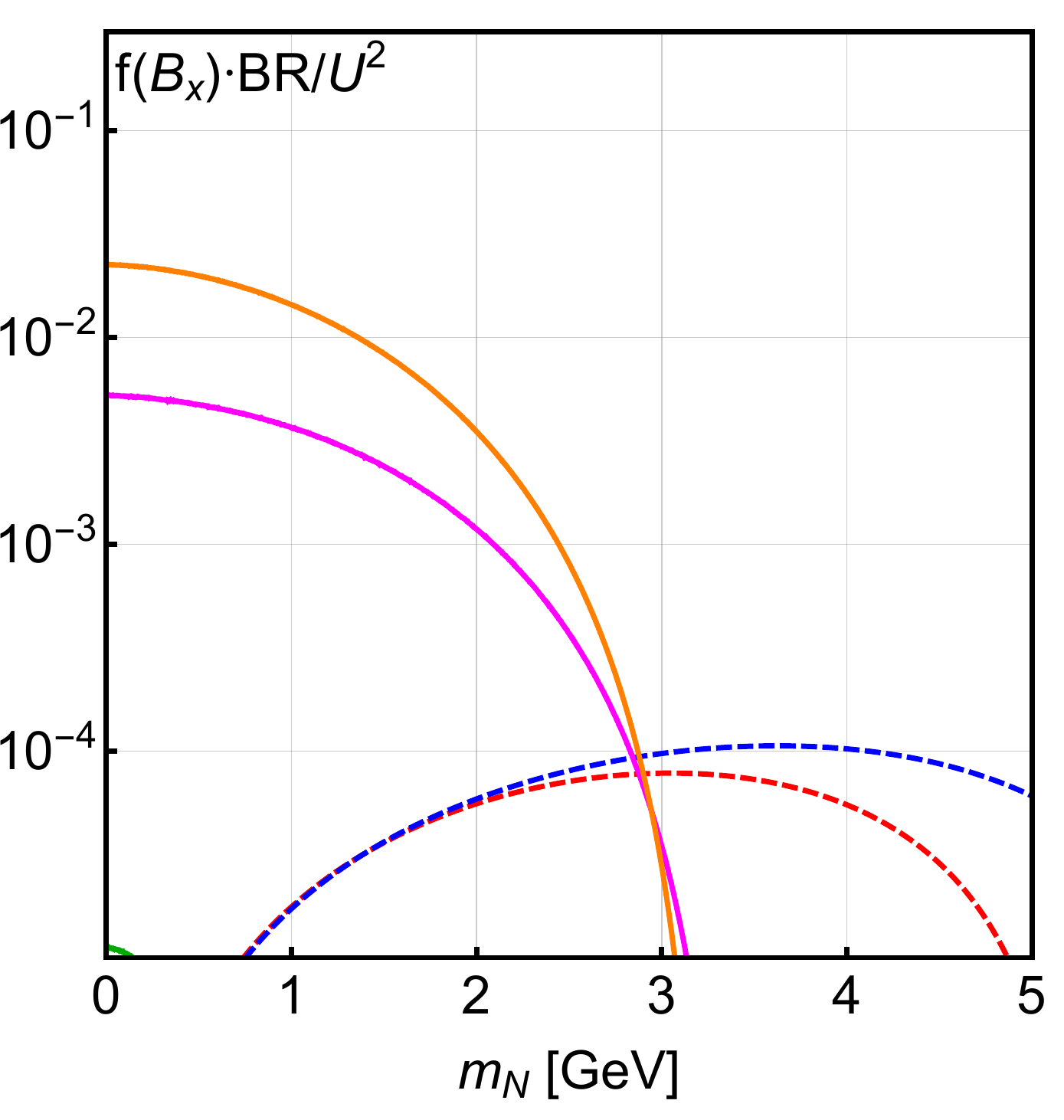}
\includegraphics[width=0.32\textwidth]{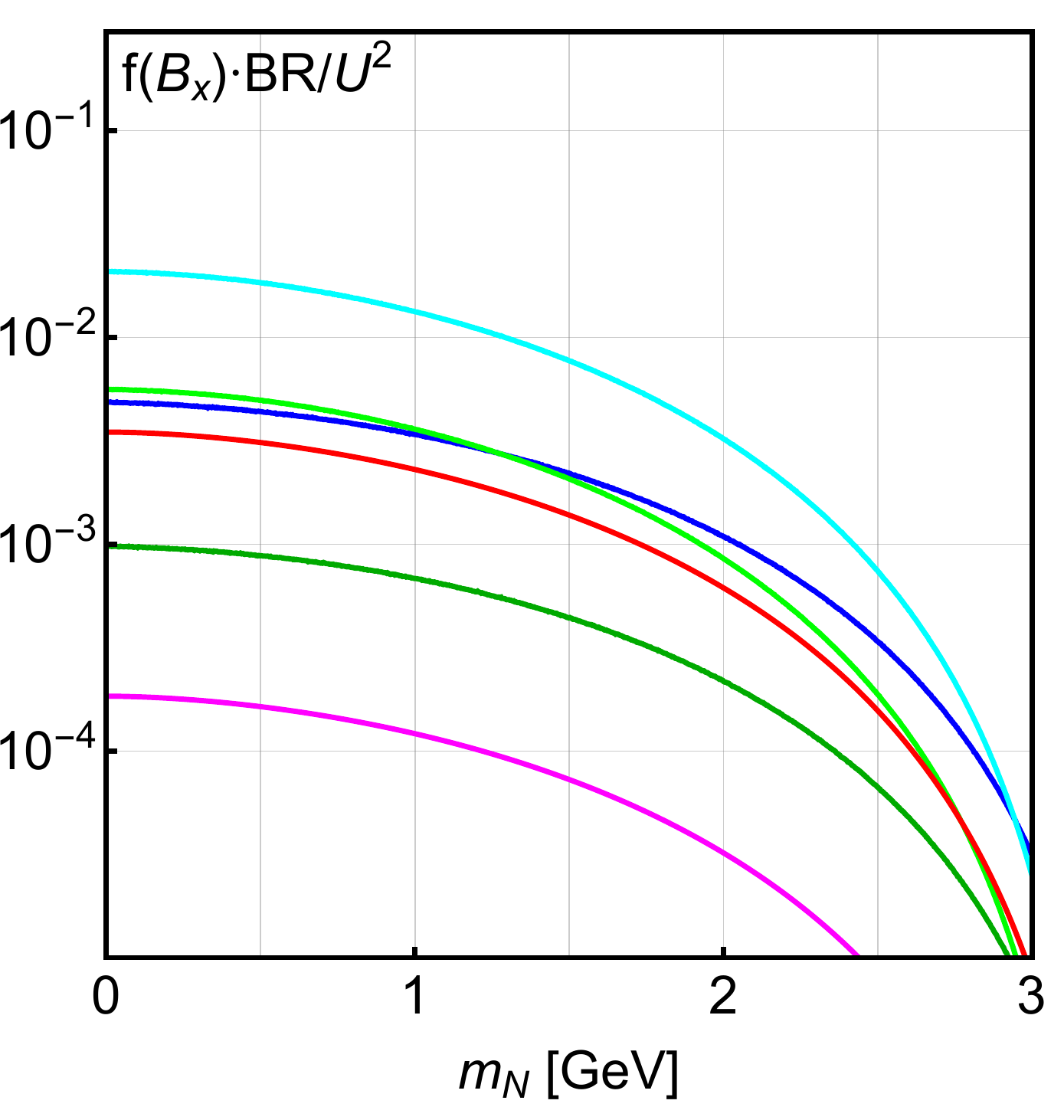}
\caption{Top: branching fractions for various $D$ and $B$-meson decay modes into HNLs as a function of the HNL mass, $m_N$, for $U_{eN}=1$ and $U_{\mu N}=U_{\tau N}=0$ following~\cite{Gorbunov:2007ak, Graverini:2133817}. Bottom: above branching fractions combined with the corresponding fragmentation fractions into different mesons.}
\label{fig:productionmodes}
\end{figure}

The list of the decay channels into sterile neutrinos that we take into account can be found in the top panels of \Figref{productionmodes} where we show relevant branching fractions~\cite{Gorbunov:2007ak} as a function of $m_N$ for $U_{eN}=1$ and $U_{\mu N}=U_{\tau N}=0$. This is then combined with the fragmentation fractions in the bottom panels of \Figref{productionmodes}. Below the kaon threshold, leptonic $K^\pm\rightarrow\,e^\mp N$ and semileptonic $K^0_{S,L}\rightarrow \pi^\pm\,e^\mp N$ decays, which are not shown in \Figref{productionmodes}, play a dominant role. In the case of nonzero mixing with the muon or tau neutrino, the corresponding SM leptons are allowed in the final states when kinematically not forbidden.

For $m_D>m_N>m_K$, decays of $D$ mesons give a dominant contribution as they are produced more abundantly at the LHC than $B$ mesons. As can be seen, the dominant production mode for HNLs is then via leptonic decays of $D_s^+$ since this process is not CKM suppressed ($V_{cs}\sim 1$). For the same reason, non-negligible contribution also comes from semileptonic decays of $D$ mesons into kaons, but these processes are more phase-space suppressed with respect to two-body decays of $D_s^+$.

In the case of $B$ mesons, which becomes important for $m_N>m_D$, the dominant contribution comes from the least CKM suppressed ($V_{cb}\sim 0.04$) channels, i.e., semileptonic decays of $B^{\pm,0}$ 
to $D$ mesons and leptonic decays of $B_c^\pm$. For $m_N>m_B-m_D$, we add important contributions from leptonic decays of $B^{\pm}$ that are more strongly CKM suppressed than decays of $B_c^\pm$, but are less sensitive to a detailed modeling of fragmentation into $B_c^\pm$. 

If sterile neutrinos mix dominantly with the tau neutrino, HNL production in decay channels of $D$ mesons is suppressed for $m_N>m_D-m_\tau$. In this case, provided $m_N<m_\tau$, tau decays give a dominant contribution to HNL production, where tau leptons typically originate from $D_s \to \tau \nu_\tau$ decays.

The kinematic distributions for HNLs produced in $B^\pm$($D_s^\pm$)-meson decays are shown in the top central (bottom central) panels of \Figref{PvsT} for $m_N=3~\gev$ and $U_{eN} = 2\times 10^{-3}$ ($m_N=1~\gev$ and $U_{eN} = 4\times 10^{-4}$) while other mixing angles are set to zero. The choice of the mixing angles for both points is dictated by the current bounds that exclude values above ${\rm a\ few}\times 10^{-3}$ or  ${\rm a\ few}\times 10^{-4}$ for $m_N$ above or below the $D$-meson threshold, respectively.
For an integrated luminosity of $3~\iab$ at the $13~\tev$ LHC one can obtain about $10^6$ ($10^8$) sterile neutrinos with these masses and mixing angles.

%%%%%%%%%%%%%%%%%%%%%%%%%%%%%
\subsection{HNL decays in FASER}

\begin{figure}[t]
\centering
\includegraphics[width=0.48\textwidth]{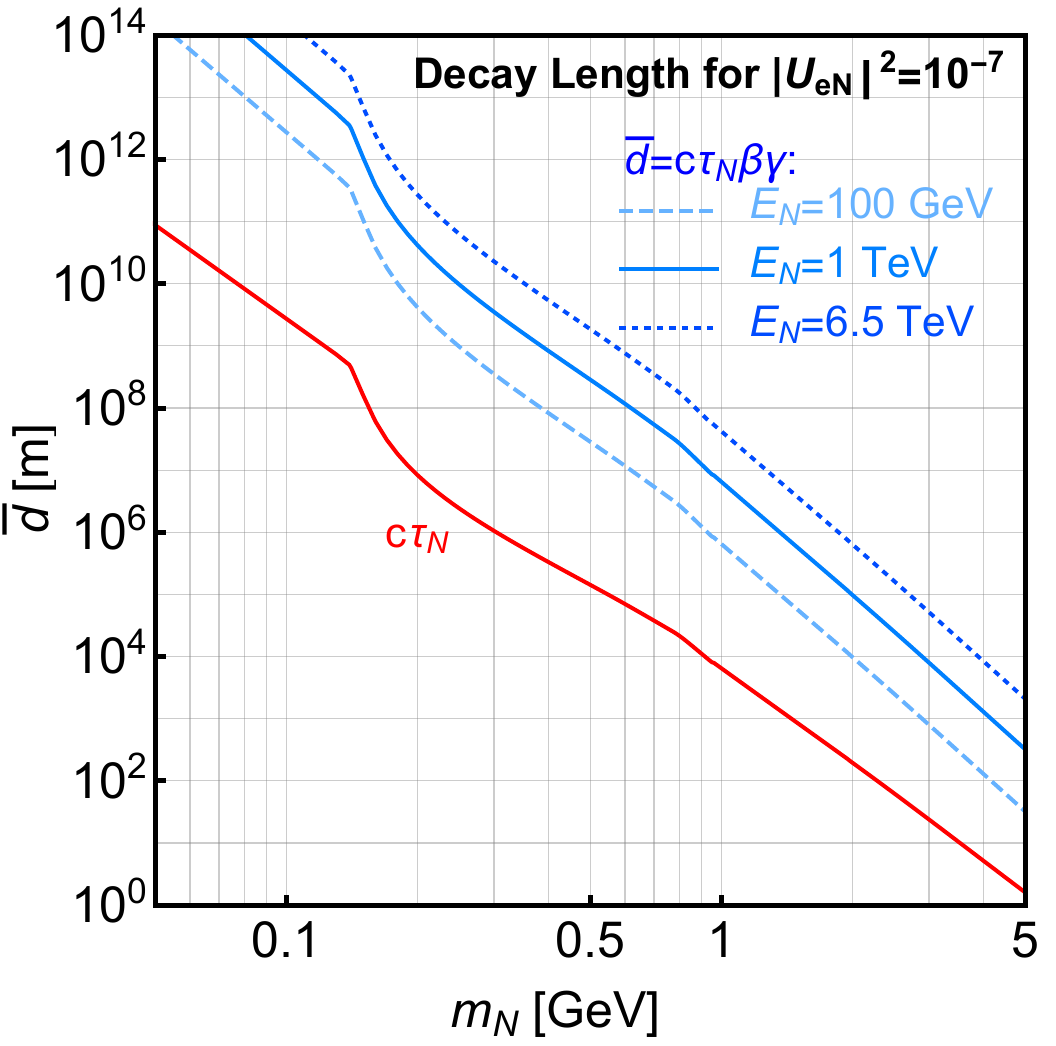}
\includegraphics[width=0.48\textwidth]{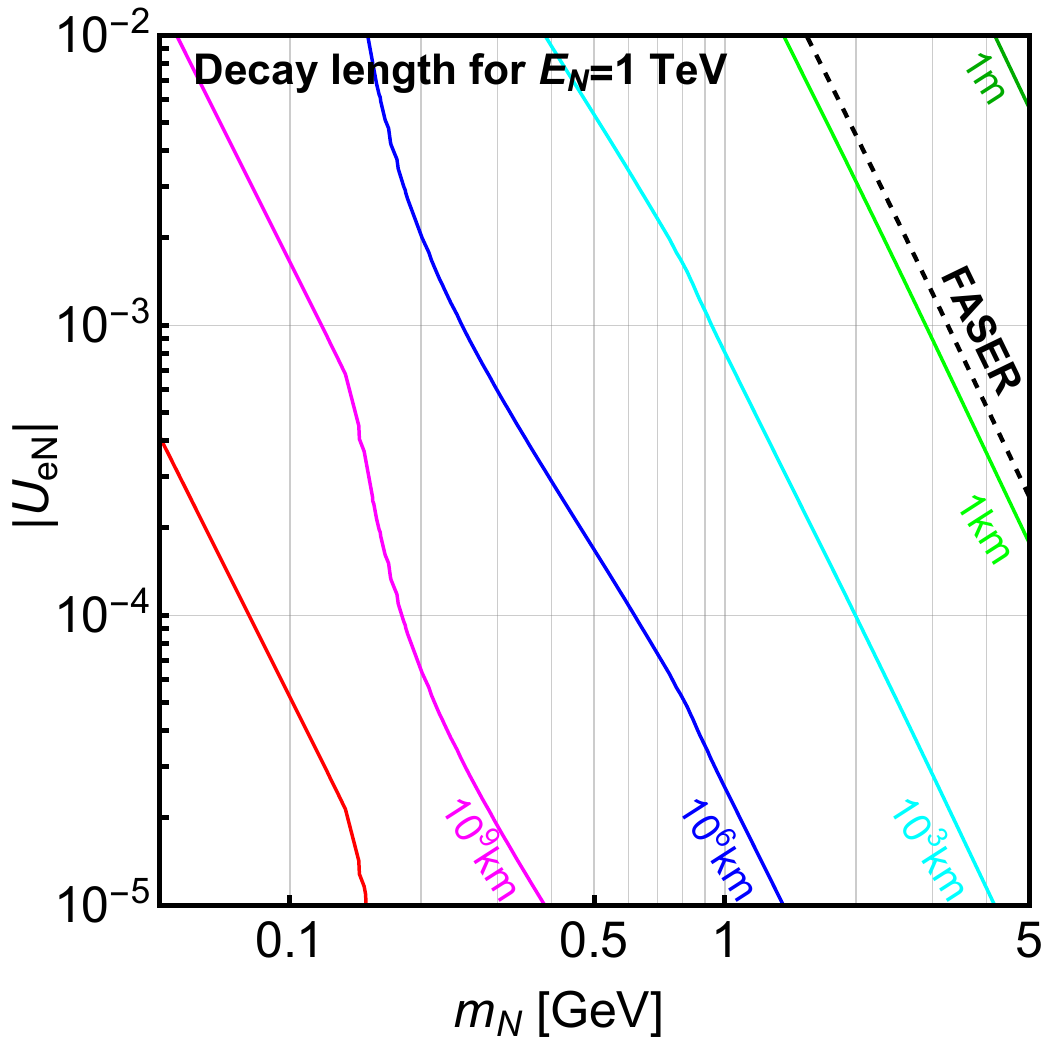}
\caption{Left: decay length $c\tau$ (solid red line) of HNLs with $|U_{eN}|^2=10^{-7}$ as a function of the HNL mass, $m_N$. Dashed (solid, dotted) blue lines correspond to the value of $\bar{d}=\gamma\,\beta\,c\,\tau$ of boosted HNL with $E = 100~\gev$ ($1~\tev$, $6.5~\tev$). Right: contours of constant decay length of HNL with $E_N=1~\tev$ in the $(m_N,U_{eN})$ plane.  The black dashed line corresponds to the distance to the FASER detector $L_{\textrm{max}}=480~\m$.}
\label{fig:lifetime}
\end{figure}

Once produced in the forward direction, HNLs can travel long distances before decaying as their decay width is suppressed by the square of the mixing angle, $|U_{\ell N}|^2$. In the left panel of \Figref{lifetime} we show typical decay lengths $c\tau_N$, where $\tau_N$ is the sterile neutrino lifetime, assuming nonzero mixing with the electron neutrino with $|U_{eN}|^2=10^{-7}$. In our estimates, we follow~\cite{Helo:2010cw} (see also~\cite{Cvetic:2010rw,Dib:2014iga,Cvetic:2015naa} for a related discussion) in which the HNL decays are treated in a channel-by-channel approach up to the mass of $\eta'$ meson, while above this threshold the inclusive approach is employed. The resulting HNL lifetime may differ by a factor of a few for $m_N\simeq m_B$ in comparison with a channel-by-channel calculation extended up to the larger masses~\cite{Atre:2009rg}, while the differences are smaller for lighter $N$. 
 
The lifetime in case of mixing with the muon or tau neutrino is of similar order as somewhat more long-lived HNLs obtained in the latter scenario due to kinematical suppression of modes with the tau lepton in the final state. Importantly,  as shown in the central panels of \Figref{PvsT}, HNLs produced at the LHC in the forward direction are typically boosted, which further increases their decay length in the lab frame. As can be seen in \Figref{lifetime}, it often exceeds $1~\km$ with lower values possible only for $m_N\gtrsim m_D$. Hence, only a small fraction of produced HNLs effectively decay at the location of FASER. However, as we will see below, even a small detector might find enough of HNL decays to probe the interesting part of the parameter space of this model. 

The probability of decay within the detector volume is given by
\begin{equation}
\mathcal{P}^{\rm det}_N = \left(e^{-L_{\rm min}/\bar{d}} - e^{-L_{\rm max}/\bar{d}}\right)\Theta\left(R-\tan\theta_N\,L_{\rm max}\right)\simeq (\Delta/\bar{d})\,\Theta\left(R-\tan\theta_N\,L_{\rm max}\right),
\label{eq:probability}
\end{equation}
where $\bar{d} = \gamma\,\beta\,c\,\tau_N$, $\theta_N$ is the angle between the sterile neutrino momentum and the beam axis in lab frame, $R$ is the detector radius, $L_{\rm min}$ corresponds to the position of the detector and $L_{\rm max} = L_{\rm min}+\Delta$ where $\Delta$ is the length of FASER along the beam axis. Since typically $\bar{d}\gg L_{\rm max}$, the probability of decay within the detector volume scales linearly with $\Delta$ and is inversely proportional to the decay length $\bar{d}$ as shown in \eqref{eq:probability}. The simple \eqref{eq:probability} is modified in case the particles decaying into HNLs have non-negligible lifetimes and, therefore, can travel sizable distances before decaying. In particular, it is true for $K^0_L$ and $K^\pm$ decays into HNLs. The former is required to decay before hitting the first neutral absorber, while the latter is required before reaching the first magnets deflecting their trajectories.

We consider two representative detector setups at the far location:
\begin{equation}
L_{\rm max} = 480~\m,\hspace{0.5cm}\Delta = 10~\m,\hspace{0.5cm}R = 20~\cm, 1~\m.
\label{eq:detector}
\end{equation}
A small detector with $R=20~\cm$ is technologically more appealing and is large enough to achieve the full reach of FASER in the case of search for dark photon~\cite{Feng:2017uoz} and axion-like particles~\cite{FASERALPs}. On the other hand, larger detector with $R=1~\m$ is desired if new physics particles are produced dominantly in heavy-meson decays like, e.g., dark Higgs bosons~\cite{Feng:2017vli}.

In the right panels of \Figref{PvsT} we show the kinematical distribution  of sterile neutrinos that decay within the volume of FASER for the same points in the model parameter space as mentioned above. As can be seen in the top panel, in case of $m_N=3~\gev$, only high-energy HNLs with $E_N\gtrsim$ few hundred \gev\ can reach the detector. On the other hand, as illustrated in the bottom panel, HNLs with lower masses and couplings have larger lifetime and therefore can survive until they reach FASER even if they have energies below $100~\gev$.

An approximate region of the parameter space in the $(m_N,U_{eN})$ plane that can be probed by FASER can be deduced from the right panel of \Figref{lifetime} where we show the contours of constant decay length of  sterile neutrinos with $1~\tev$ energy. We can see that most parts of the parameter space correspond to the long lifetime limit $\bar{d} \gg L_{\rm max}$ in which the HNL tends to overshoot the detector and the event rate is limited by the detector length. On the other hand, at large masses $m_N > m_D$ and mixings $U_{eN}\sim 10^{-2}$ the decay length becomes short, $\bar{d} < L_{\rm max}$, and only highly boosted particles can reach FASER.  We illustrate this in the right panel of \Figref{lifetime} where the distance to the FASER detector is explicitly marked with the black dashed line.

\begin{figure}[t]
\centering
\includegraphics[width=0.32\textwidth]{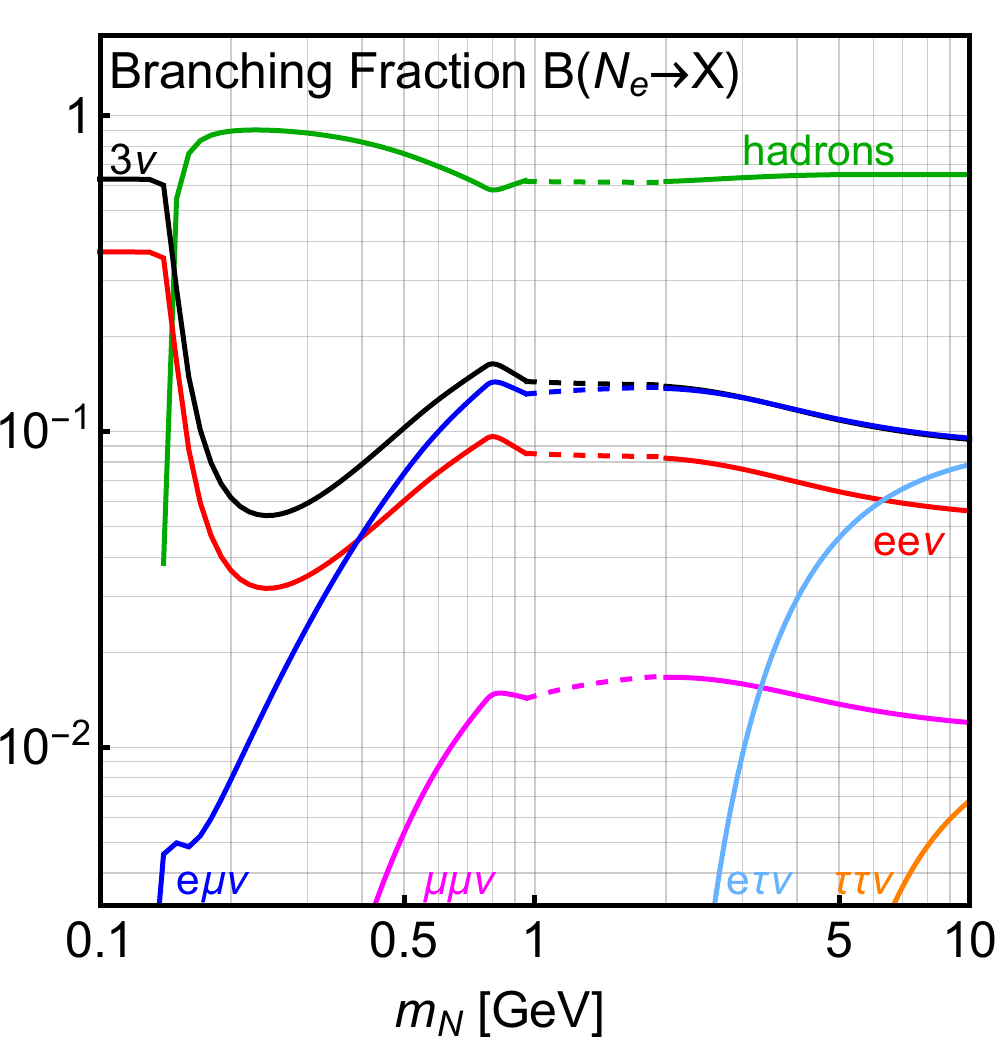}
\includegraphics[width=0.32\textwidth]{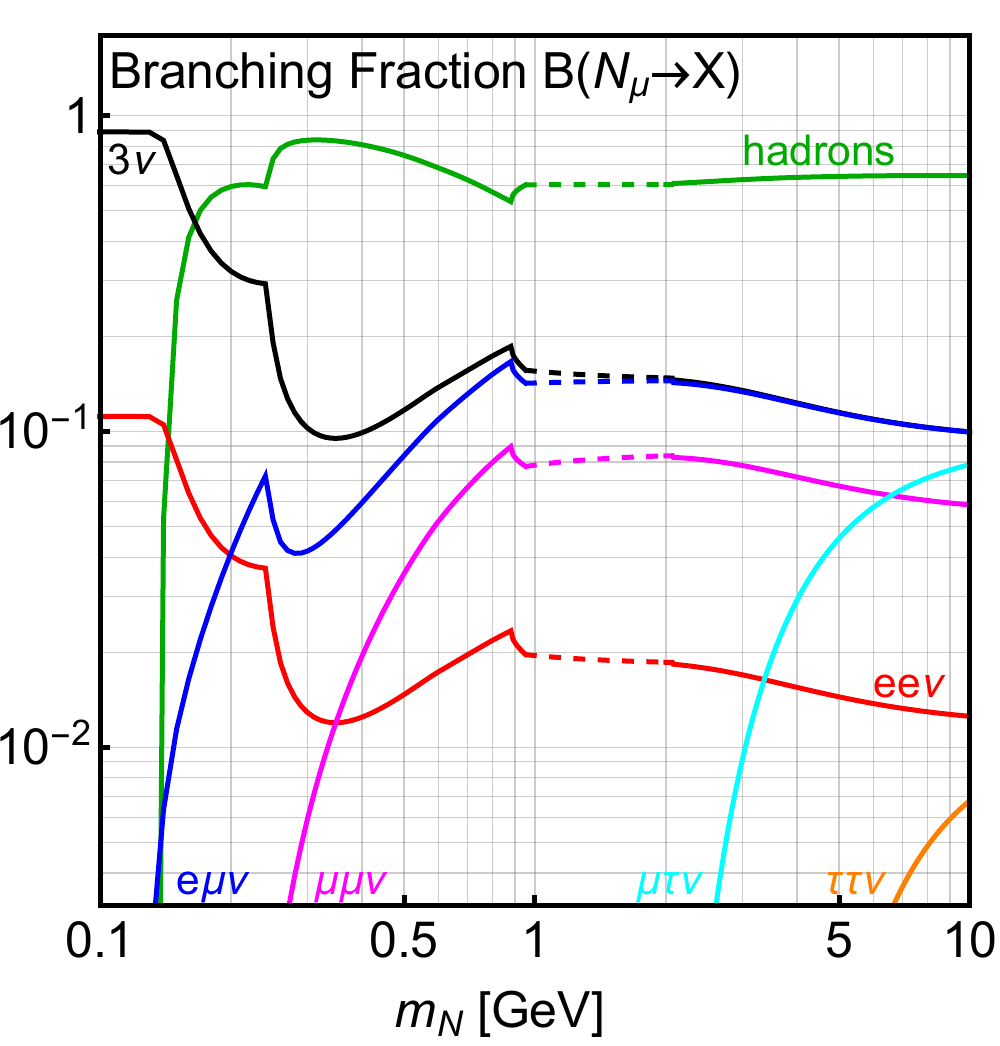}
\includegraphics[width=0.32\textwidth]{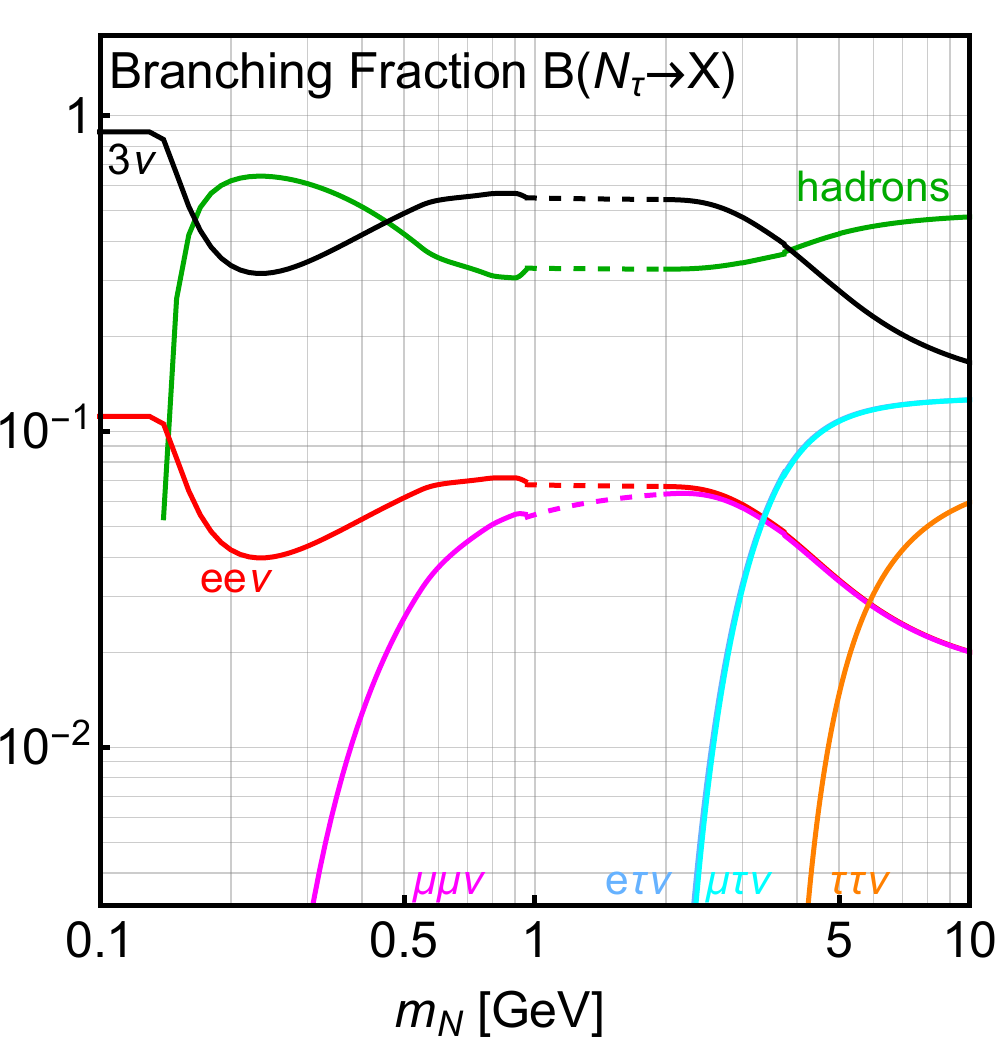}
\caption{Dominant branching fractions as a function of the sterile neutrino mass, $m_N$, for nonzero $U_{eN}$ (left), $U_{\mu N}$ (center) and $U_{\tau N}$ (right). See the text for details.}
\label{fig:Bfractions}
\end{figure}

In \Figref{Bfractions} we show the main decay channels of sterile neutrinos for each of the three scenarios with nonzero mixing angles that we consider. The leptonic channels include various $\nu\,l^+\,l^-$ final states, as well as invisible decay modes into three SM neutrinos. For $m_N>m_\pi$, various two-body decays into a pair of a lepton and a meson are possible which we implement following~\cite{Gorbunov:2007ak, Atre:2009rg,Helo:2010cw}. In particular, the green lines for $m_N\lesssim m_{\eta'}\simeq 957.8~\mev$ correspond to combined branching fraction to various light unflavored mesons including $\pi^{0,\pm}$, $\eta$, $\rho^{0,\pm}$, and $\omega$, as well as to strange mesons, $K^{(\ast),0,\pm}$. As can be seen, the branching fractions into hadrons typically dominate in the mass range of interest in scenarios with nonzero mixing with the electron and muon neutrino, while a phase-space suppression can make this contribution smaller than the invisible decay channel into three neutrinos for $N$s mixing with only tau neutrinos. Various threshold effects are visible in the lines corresponding to hadronic decay channels, \eg, when decays into pions or $\rho$ and $\omega$ mesons become kinematically allowed. We further assume, following~\cite{Helo:2010cw}, that for $m_N\gtrsim m_{\eta'}$ one can consider quarks as degrees of freedom in the final state in decays $N\rightarrow (\nu/l)q_1\bar{q}_2$. This also allows us to treat possible HNL decays with multiple mesons in the final state (for a recent discussion, see~\cite{Dib:2018iyr}). We note, however, that for values of the HNL mass in a range $m_{\eta'}\lesssim m_N\lesssim 2~\gev$, above which the use of an inclusive approach is better justified, the obtained branching fractions should be treated less robustly as denoted by the use of dashed lines in \Figref{Bfractions}. 

%%%%%%%%%%%%%%%%%%%%%%%%%%%%%
%%% Reach
%%%%%%%%%%%%%%%%%%%%%%%%%%%%%
\section{Sensitivity of FASER to HNL\lowercase{s}\label{sec:reach}}

The expected signal from sterile neutrinos decaying in FASER consists of two simultaneous high-energy charged tracks, which originate from a vertex inside the detector and the combined momentum of which points into the direction of the IP. While detailed background analysis goes beyond the scope of this paper, one can argue that natural and infrastructure-based shielding, as well as specific properties of the signal, should allow one to disentangle it from background~\cite{Feng:2017uoz}. In particular, SM particles produced abundantly at the IP would be either deflected by the LHC magnets or stopped by neutral absorbers before they reach FASER. On the other hand, energetic particles produced in beam-gas collisions in the beam pipe close to the detector would effectively lose their energy in the rock and concrete separating the side tunnel from the main tunnel. The kinematic features of the signal, such as the directionality of the charged particles observed in FASER, would provide an additional handle that could help to discriminate between these backgrounds and the signal events, as well as to reject events with cosmic-ray origin. In the following, we will for simplicity assume zero background when discussing FASER's sensitivity.

FASER's sensitivity reach can also be affected by the finite detection efficiency. In particular, we correct the expected number of events to take into account sterile neutrino decays into three SM neutrinos. In addition, the reconstruction of the direction of visible tracks in the detector might be affected by possible multibody decays of the HNLs and the presence of invisible active neutrinos in the final state. However, given large boost of the HNL and its decay products, we expect that such reconstruction should be sufficiently good to help disentangle between background and signal events. In the following, we will assume $100\%$ efficiency for other decay channels, while a more detailed analysis of this effect is postponed to a later dedicated study. 

\begin{figure}[t]
\centering
\includegraphics[width=0.48\textwidth]{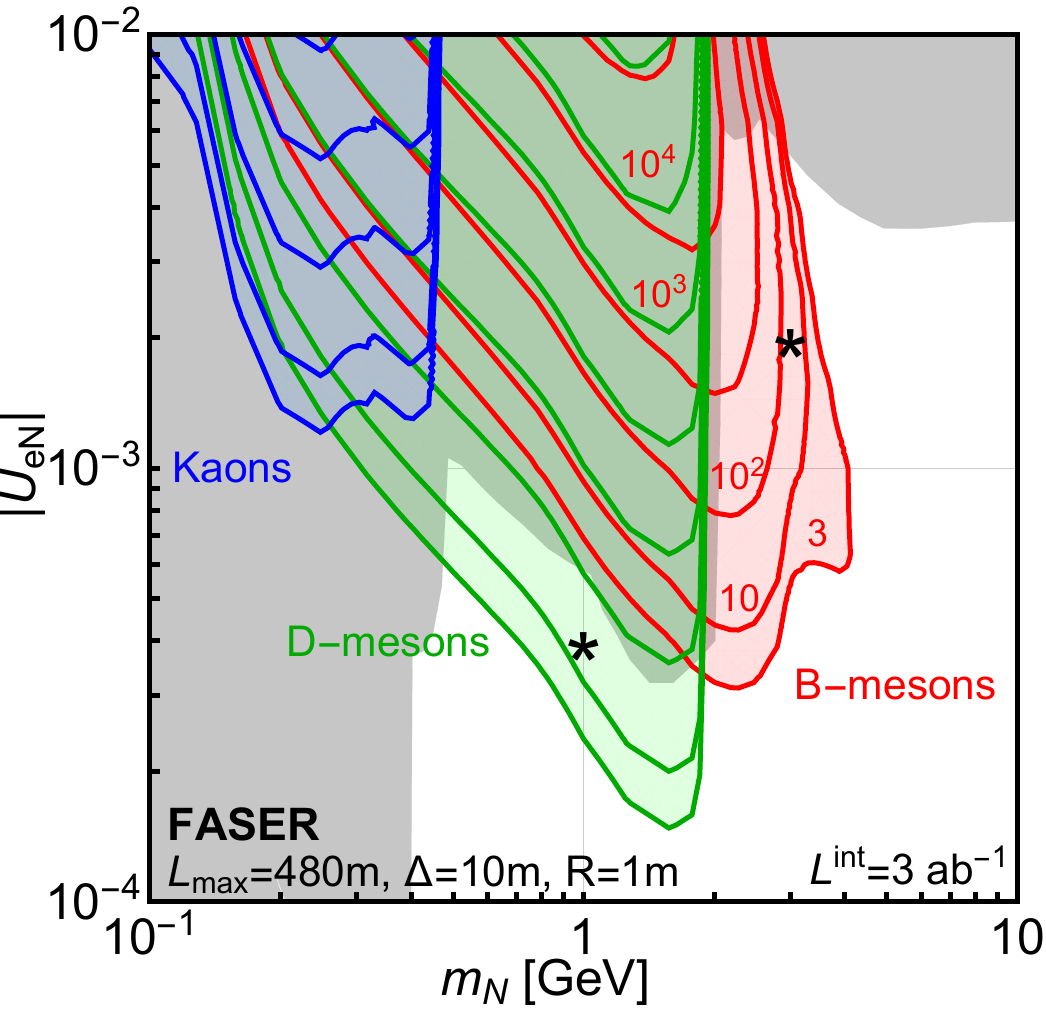}
\includegraphics[width=0.48\textwidth]{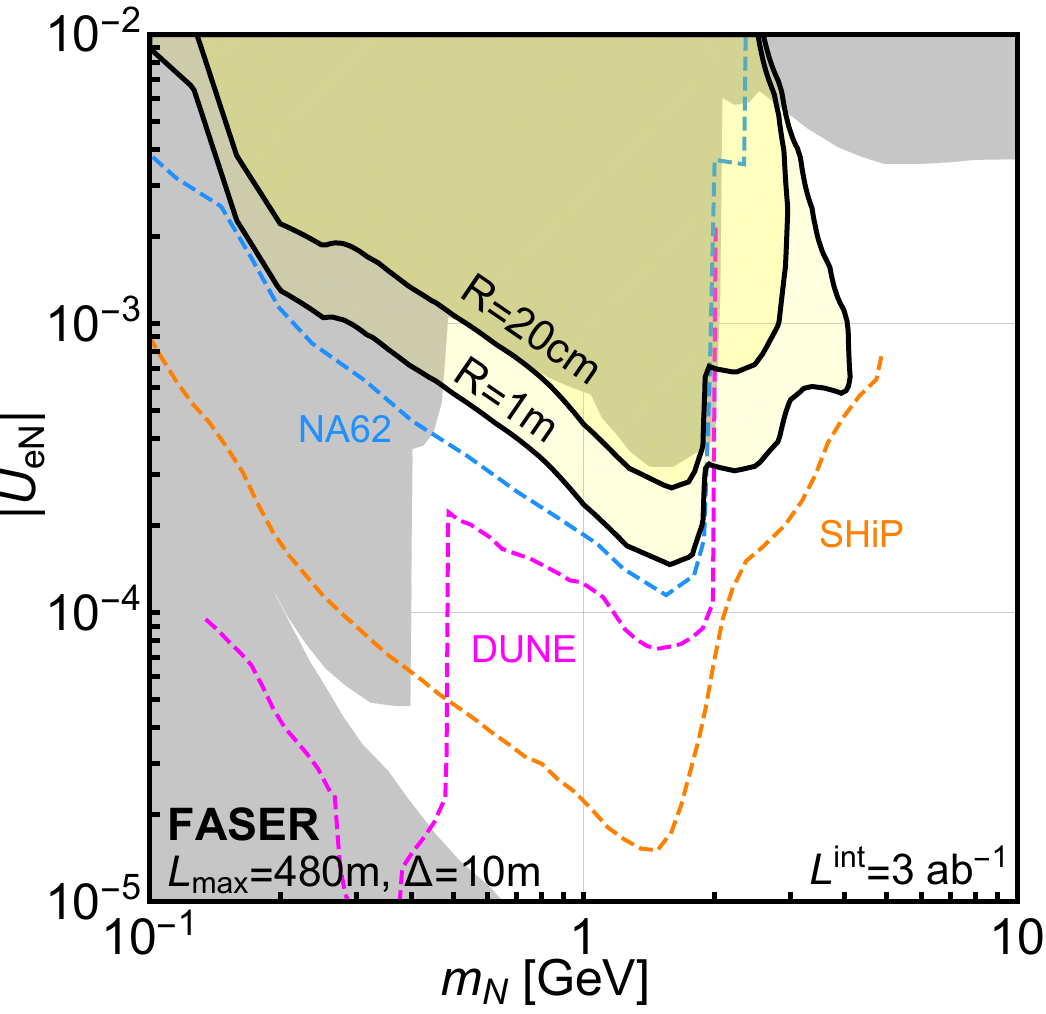}\\
\caption{Sensitivity reach of FASER to HNLs mixing with the electron neutrino in the $(m_N,U_{eN})$ plane. In the left panel, contributions from sterile neutrinos produced in kaon (blue), $D$-meson (green) and $B$-meson (red) decays are shown with the lines of a fixed number of signal events $n_{\rm sig} = 3,10,10^2,10^3,10^4$. The current exclusion bounds are shown in the gray band (see the text for details). The black stars correspond to the benchmark points for which kinematical distributions are shown in \Figref{PvsT}. The shaded areas in the right panel correspond to the reach of $R=20~\cm$ and $R=1~\m$ detectors assuming $3~\iab$ integrated luminosity. The sensitivity reach of the proposed SHiP experiment~\cite{Alekhin:2015byh} (orange dashed line), the planned DUNE experiment~\cite{Adams:2013qkq} (pink dashed line), and the NA62 experiment~\cite{Drewes:2018gkc} (blue dashed line) are shown for comparison.}
\label{fig:results}
\end{figure}

In \Figref{results} we show the sensitivity reach of FASER to search for sterile neutrinos that mix only with the electron neutrino, where the gray bands correspond to the current exclusion limits. In the left panel we show the sensitivity of $R=1~\m$ detector for scenario with $U_{eN}\neq 0$ decomposed into contributions from kaon (blue), $D$-meson (green) and $B$-meson (red) decays. Typically lighter-meson decays dominate when kinematically allowed since they are produced more abundantly at the LHC. The solid lines correspond to a fixed number of signal events, $n=3,\ 10, \ 10^2,\ 10^3,\ 10^4$. As can be seen, one can expect up to $\sim 1000$ events with HNL origin in FASER for $m_N\gtrsim m_D$ and $\sim 10$ signal events for sterile neutrinos produced in $D$-meson decays. Clearly, possible changes in the number of signal events by a factor $\mathcal{O}(1)$ from a finite efficiency of the detector or non-negligible background would have a mild impact on the reach especially in the case of $B$-meson decays. The region most sensitive to such changes corresponds to $m_N>m_B-m_D$ where dominant semileptonic decays of $B$ mesons into HNL and $D$ mesons are kinematically forbidden. Instead, a few events in this mass regime can be expected from leptonic $B^\pm$ and $B_c^\pm$ decays that are otherwise subdominant. This results in a characteristic bumplike shape of the sensitivity line.

In the right panel of \Figref{results} we compare the sensitivity of the $R=20~\cm$ and $R=1~\m$ detector for $3~\iab$ integrated luminosity. As can be seen, the reach for $R=1~\m$ is significantly improved with respect to the one obtained for a smaller detector. This is not a surprise since HNLs, similarly to dark Higgs bosons~\cite{Feng:2017vli}, are typically produced in heavy-meson decays and, therefore, they are less collimated around the beam axis than, \eg, dark photons produced in pion decays~\cite{Feng:2017uoz}. The reach could be further improved, especially for $m_N\lesssim m_D$, for an even larger radius, as can be deduced from the bottom right panel of \Figref{PvsT}.

\begin{figure}[t]
\centering
\includegraphics[width=0.48\textwidth]{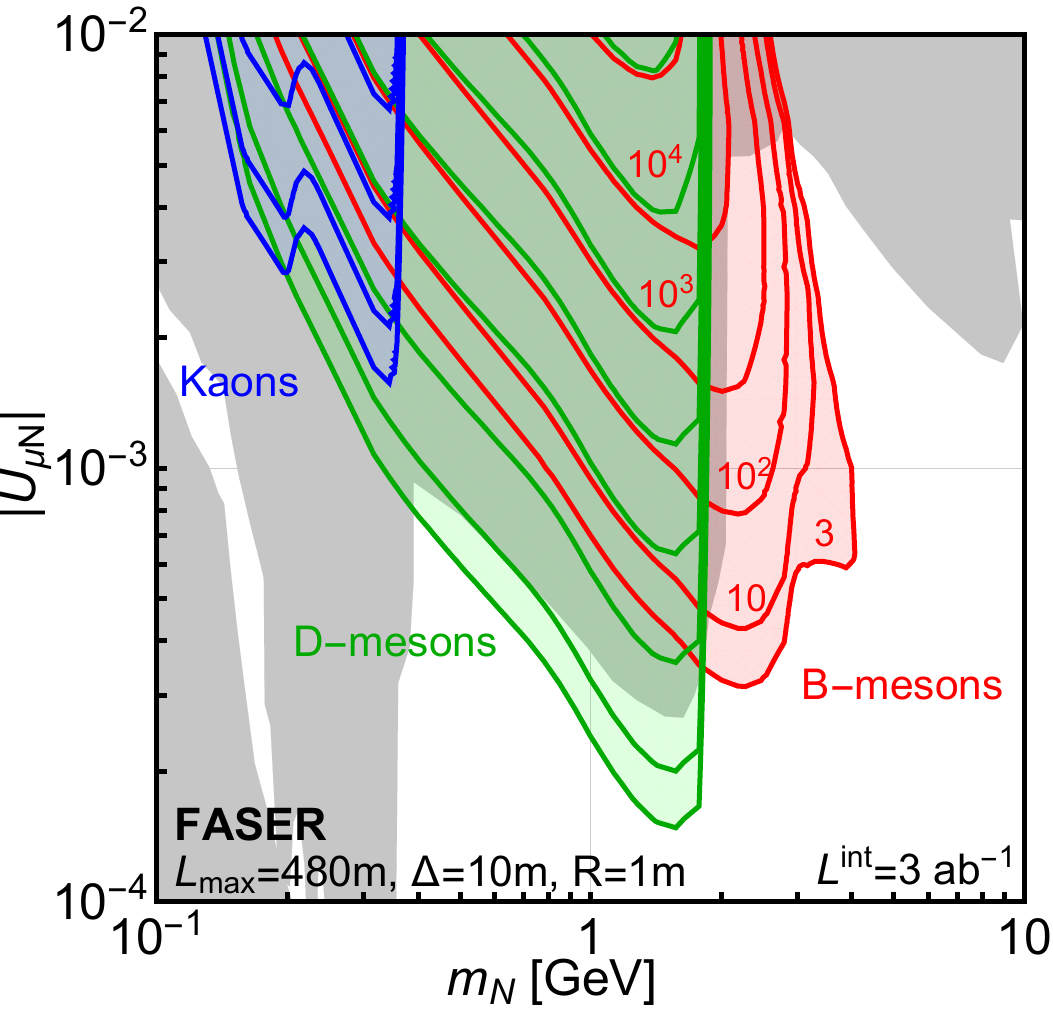}
\includegraphics[width=0.48\textwidth]{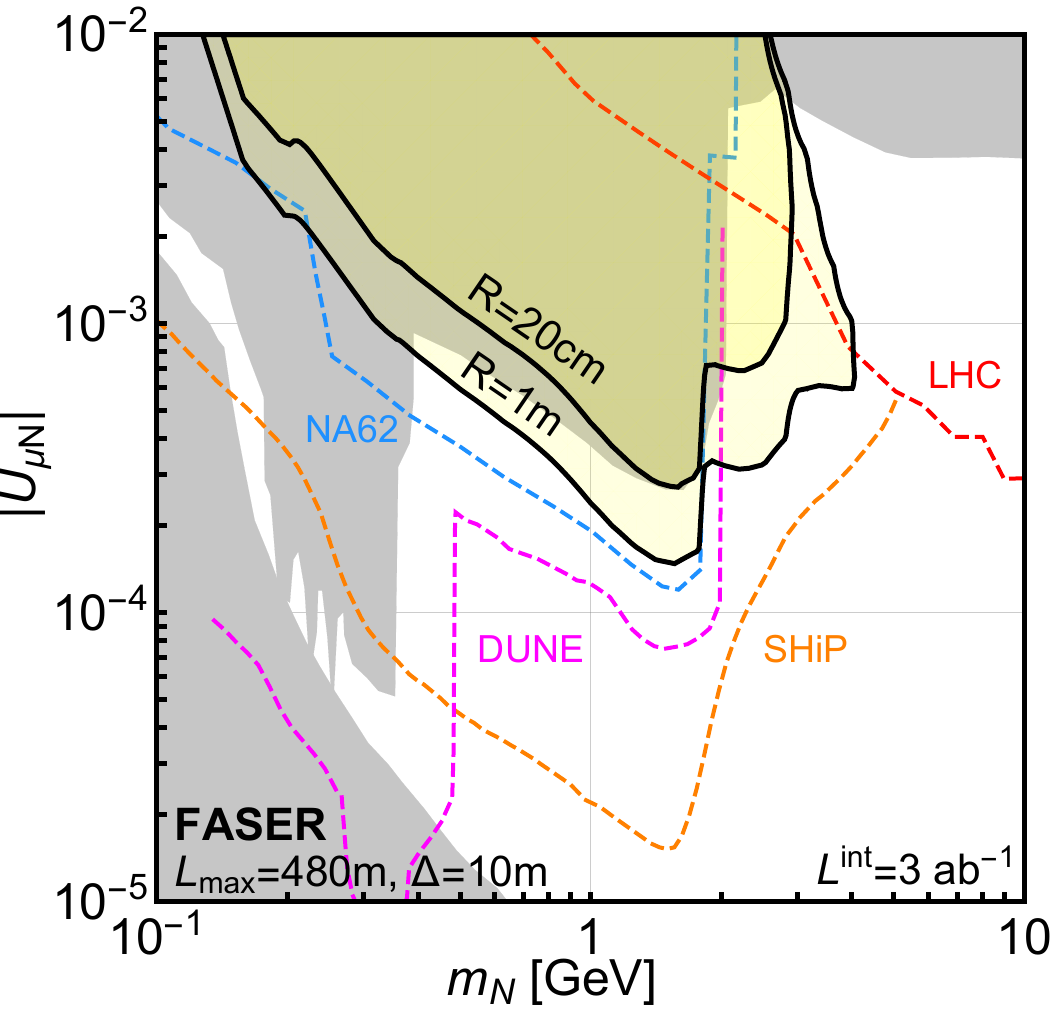}\\
\caption{Similar to \Figref{results} but for $U_{\mu N}\neq 0$ and $U_{eN}=U_{\tau N}=0$. The red line in the right panel corresponds to the LHC searches for a prompt lepton $+$ a single displaced lepton jet~\cite{Izaguirre:2015pga}.}
\label{fig:resultsmu}
\end{figure}

\begin{figure}[!h]
\centering
\includegraphics[width=0.48\textwidth]{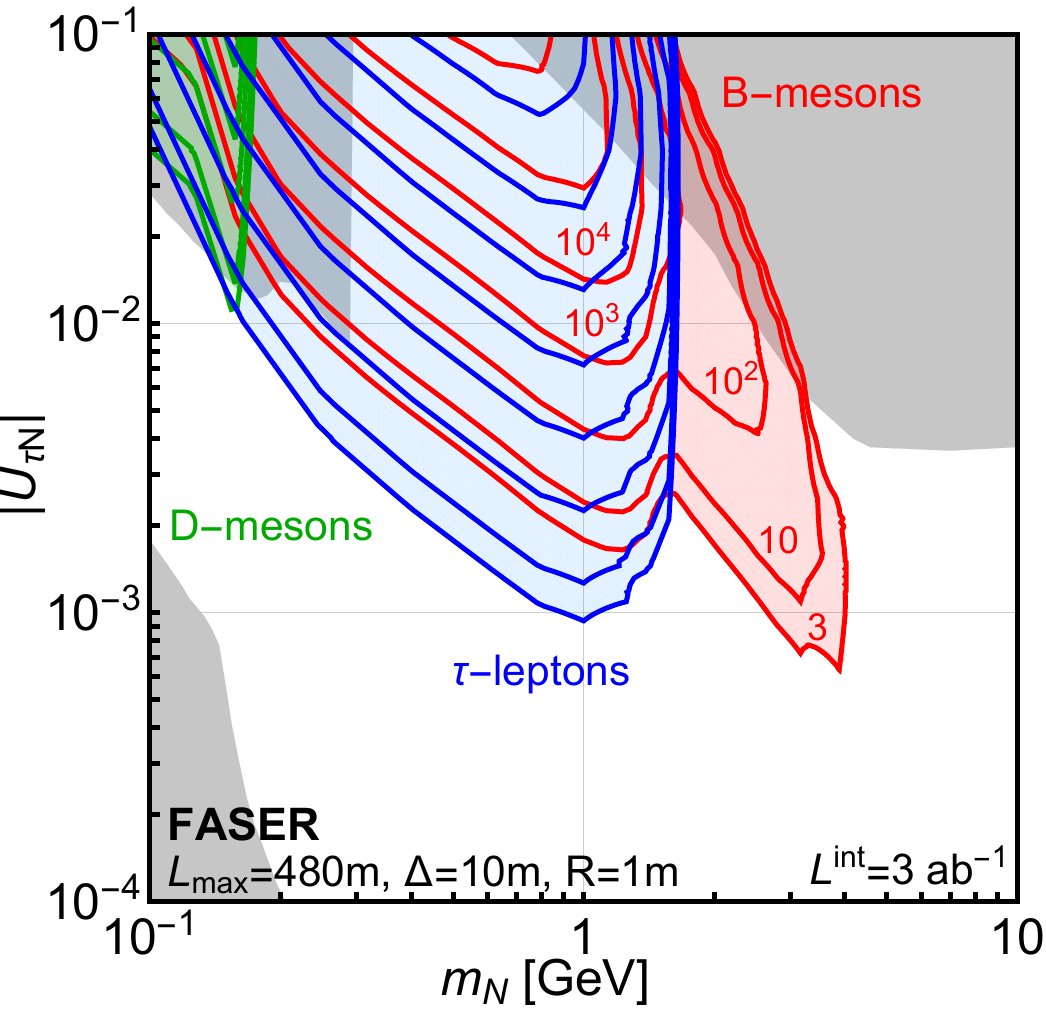}
\includegraphics[width=0.48\textwidth]{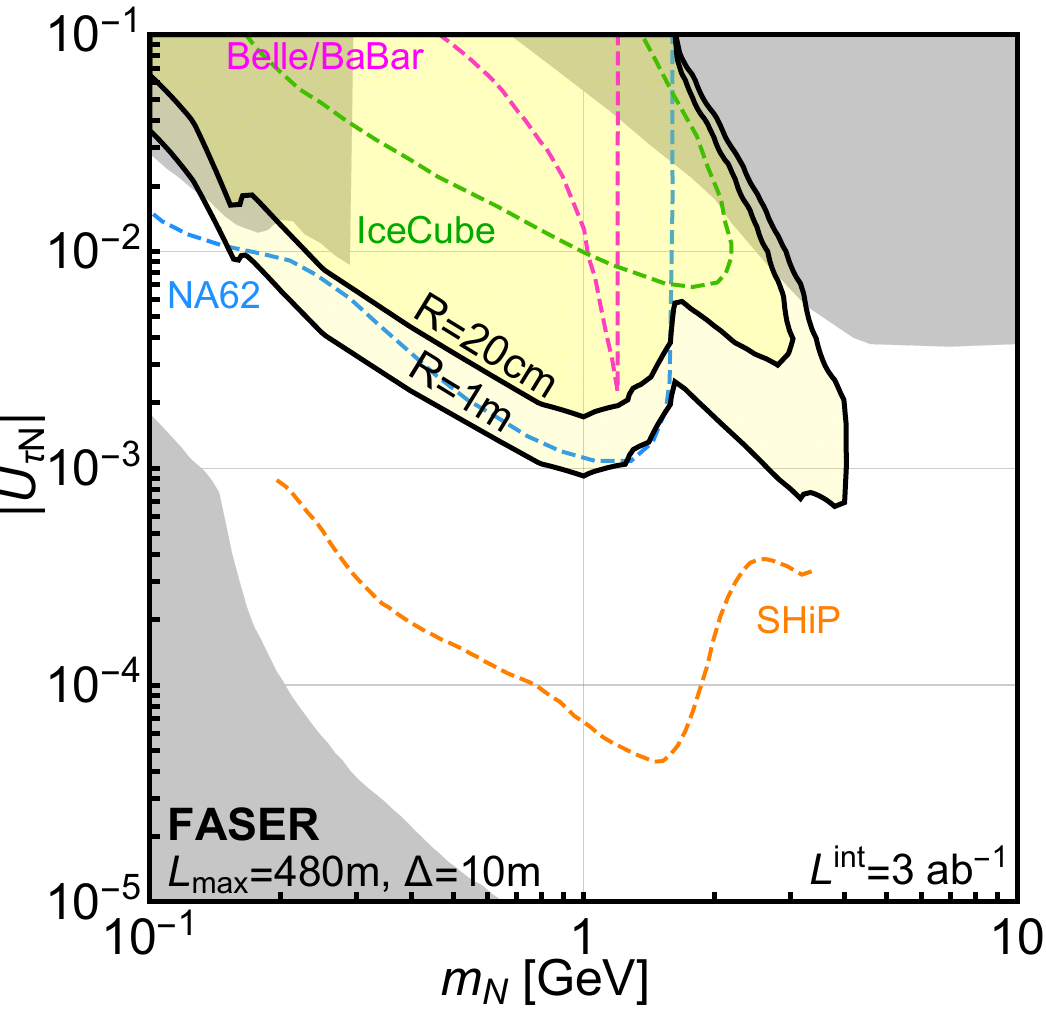}\\
\caption{Similar to \Figref{results} but for $U_{\tau N}\neq 0$ and $U_{eN}=U_{\mu N}=0$.  In the left panel the green (blue, red) shaded areas correspond to HNLs produced in decays of $D$ mesons (tau leptons, $B$ mesons). The pink dashed line in the right panel corresponds to projected sensitivity from production in $\tau$ decays in $B$ factories~\cite{Kobach:2014hea}, while the green dashed line is the search for double-cascade events in the IceCube detector~\cite{Coloma:2017ppo}.}
\label{fig:resultstau}
\end{figure}

Similarly, the cases of mixing with the muon neutrino and tau neutrino are shown in \Figref{resultsmu} and \Figref{resultstau}, respectively. As can be seen, FASER will probe parts of the yet unconstrained region of the parameter space in each of these scenarios. In the electron and muon case, current bounds below the kaon threshold are stronger than the reach of FASER. However, FASER's capability is much improved once $m_N$ grows and the typical decay length of boosted sterile neutrinos becomes comparable to the distance between the IP and the detector (compare \Figref{lifetime}). On the other hand, the scenario with nonzero mixing with the tau neutrino is much less constrained. This leads to an even larger region of the parameter space for HNLs that has not been probed yet and is characterized by an excellent discovery potential in FASER with possibly even tens of thousands of expected signal events for $3~\iab$ integrated luminosity. Importantly, the reach for $m_N> m_B-m_D-m_\tau$ corresponds to the leptonic decays of $B^\pm$ and $B_c^\pm$ and leads to a bumplike shape of the sensitivity line which is much more pronounced than in the case of nonzero mixing with the electron and muon neutrinos.

In the range of $m_N$ relevant for FASER, the most stringent bounds on sterile neutrino mixing with the electron neutrino or muon neutrino come from past and present beam-dump experiments at CERN (PS191~\cite{Bernardi:1987ek}, CHARM~\cite{Bergsma:1985is} and NA62~\cite{CortinaGil:2017mqf}) and IHEP-JINR~\cite{Baranov:1992vq}. For $m_N\gtrsim 2~\gev$, the strongest limits come from the search for $B$ decays into HNLs at Belle II~\cite{Liventsev:2013zz} and from limits on the $Z$ boson decays into HNLs from the LEP data collected by the DELPHI Collaboration~\cite{Abreu:1996pa}. For mixing with the muon neutrino other important bounds come from search for a double-peak structure in $K\rightarrow\mu\nu$ decays~\cite{Artamonov:2014urb} and the NuTeV beam-dump experiment~\cite{Vaitaitis:1999wq}. In the case of mixing with the tau neutrino, current limits are much weaker with the leading bounds coming from the CHARM~\cite{Orloff:2002de} and DELPHI~\cite{Abreu:1996pa} collaborations. If the mixing angles become low enough, strong bounds from BBN~\cite{Dolgov:2000jw,Dolgov:2000pj,Ruchayskiy:2012si} constrain the parameter space of HNLs from below.

In the scenario with nonzero mixing with the electron neutrino, $U_{eN}\neq 0$, other important bounds come from null searches of the neutrinoless double-beta decay, denoted as $0\nu\beta\beta$. The most stringent current limit on the $0\nu\beta\beta$ decay half-life $T_{1/2}^{0\nu}\gtrsim 1.07\times 10^{26}~\yr$ comes from a combined analysis of the phase-I and phase-II data acquired by the KamLAND-Zen experiment~\cite{KamLAND-Zen:2016pfg}. This can be translated (see, \eg,~\cite{Benes:2005hn,Helo:2010cw}) into an approximate limit on the mixing angle $|U_{eN}|^2/m_N \lesssim 2.1\times 10^{-8}~\gev^{-1}$. However, as discussed above, more than one sterile neutrino is required for the seesaw mechanism to generate correct active neutrino masses. In addition, some of the sterile neutrinos can be lighter than the typical momentum transfer $q\sim 100~\mev$ in the $0\nu\beta\beta$ process provided they interact weakly enough so as to not alter BBN. In this case, additional cancellations between various contributions to the $0\nu\beta\beta$ rate may occur and effectively weaken the corresponding bound~\cite{Blennow:2010th}. It is therefore not shown in \Figref{results}.

For comparison, in \Figref{results}, we also show the expected sensitivities for sterile neutrino searches in the proposed SHiP detector~\cite{Alekhin:2015byh}, the DUNE~\cite{Adams:2013qkq} and NA62~\cite{Drewes:2018gkc} experiments and the LHC searches for a prompt lepton plus a single displaced lepton jet~\cite{Izaguirre:2015pga} (see also~\cite{Antusch:2017hhu} for a relevant study for the LHCb). As can be seen, FASER has comparable sensitivity to NA62 for sterile neutrinos produced in $D$-meson decays. This region in the parameter space for both electron and muon mixing will also be probed by the future DUNE facility for which we show results following~\cite{Alekhin:2015byh}. They  correspond to the analysis performed for the five years of data-taking by the $30~\m$ long LBNE near detector with $5\times 10^{21}$ protons on target and assuming a normal hierarchy of neutrinos. 

Above the $D$-meson threshold, the abundant production of forward $B$ mesons at high-energy $pp$ collisions at the LHC works in favor of FASER. This allows FASER to probe the region between the aforementioned planned searches and the expected reach of the LHC searches~\cite{Izaguirre:2015pga} as shown for the scenario with $U_{\mu N}\neq 0$ in the left panel of \Figref{resultsmu}. The LHC searches will also be sensitive to other scenarios with nonzero $U_{eN}$ or $U_{\tau N}$. However, as discussed in~\cite{Izaguirre:2015pga}, dedicated analysis of relevant backgrounds and tagging efficiency would have to be performed by experimental collaborations to obtain solid results. Therefore, they are not shown in \figsref{results}{resultstau}, where we present the reach for the $U_{eN}\neq 0$ and $U_{\tau N}\neq 0$ cases, respectively. 

In case of nonzero mixing with the tau neutrino, $U_{\tau N} \neq0$, HNL production in $D$-meson decays is restricted to $m_N < m_D - m_\tau$. The main contribution comes from the decay $D_s^+ \to \tau N$. 
Larger masses $m_D - m_\tau < m_N \lesssim m_\tau$ can be probed by hadronic and leptonic tau decays, $\tau \to N \pi, N K^+ , N \rho^+ $ and $\tau \to N \ell \nu_\ell$. Here the $\tau$s are mainly produced via the decay $D_s^+ \to \tau \nu$. This region of the parameter space can also be partially probed by searches based on $\tau$ production in $B$ factories~\cite{Kobach:2014hea} and their subsequent decays into sterile neutrinos, as 
well as by search for double-cascade events initiated by high-energy neutrinos up-scattering into sterile neutrinos in the IceCube detector~\cite{Coloma:2017ppo}.

The strongest expected reach in the range of the HNL masses probed by FASER could be obtained by the proposed SHiP detector. In order to compare the reach of FASER to the one of SHiP, it is useful to note that to a good approximation the number of expected events scales like
\begin{equation}
N_{\rm ev} \sim N_{\rm HNL}\times \mathcal{P}^{\rm det}_N \sim |U_{\ell N}|^4 \times N_{\rm mes}\times \frac{\Delta}{E_{\rm mes}},
\label{eq:numbevest}
\end{equation}
where in the second step we used \eqref{eq:probability} and for simplicity we neglected the geometrical acceptance of the detector. The expected number of mesons produced for a given experiment is denoted by $N_{\rm mes}$, while $E_{\rm mes}$ is the typical energy of mesons that give rise to sterile neutrinos decaying within the volume of each detector. For both considered experiments the relevant numbers read
\begin{equation}
\begin{aligned}
\textbf{SHiP:}&     
&N_{D}&\sim 3.4\times 10^{17},    
&N_{B}&\sim 3.2\times 10^{13}, 
&E_{\rm mes}&\sim 25~\gev,
&\Delta &\sim 50~\m,\\
\textbf{FASER:}& 
&N_{D}&\sim 1.9\times 10^{16},     
&N_{B}&\sim 1.4\times 10^{15}, 
&E_{\rm mes}&\sim 1 ~\tev, 
&\Delta &\sim 10~\m
\label{eq:detdata}
\end{aligned}
\end{equation}
Combining \eqref{eq:numbevest} and \eqref{eq:detdata}, one obtains for SHiP  a number of events with $B$-meson origin only about five times larger. Given that in the regime of large lifetime the number of expected events scales like $N_{\rm ev}\sim |U_{\ell N}|^4$, one obtains only slightly better sensitivity (by a factor of $\lesssim 2$) of SHiP to sterile neutrinos produced in $B$-meson decays in the $(m_N,U_{\ell N})$ plane. In addition, for somewhat heavier $B_c^\pm$ mesons, FASER can gain even more in meson production from much larger collision energy than SHiP. As a result, the sensitivity of both detectors is comparable at the largest accessible values of $m_N\sim 4~\gev$. The loss of sensitivity for these large masses and increasing mixing angles to values $U_{\ell N}\gtrsim 10^{-3}$ can be understood since then the sterile neutrino lifetime becomes too low for HNLs to reach FASER.

On the other hand, production of $D$ mesons is less suppressed for the SHiP center-of-mass energy $27~\gev$, and therefore larger luminosity allows one to obtain $\sim 10^3-10^4$ more HNL events with $D$-meson origin in SHiP than in FASER. Hence, the reach in the mixing angle is worse for FASER by about an order magnitude as can be seen in \Figref{results}. This is also true for sterile neutrinos produced in $\tau$ decays, as seen in \Figref{resultstau}, since they typically come from $D$-meson decays.

%%%%%%%%%%%%%%%%%%%%%%%%%%%%%
%%% Additional Models
%%%%%%%%%%%%%%%%%%%%%%%%%%%%%

\section{Beyond minimal seesaw\label{sec:beyond}}

In previous sections, we have analyzed in details FASER's sensitivity reach to HNLs described effectively by only their mass and the mixing angles with the active neutrinos. We will now discuss how this search can be relevant to selected more complex scenarios going beyond the minimal seesaw mechanism with only right-handed neutrinos added to the SM. In particular, the reach for historically very important left-right symmetric models is discussed in \secref{LR}. In \secref{mirror}, we illustrate the possible connection of the sterile neutrino search in FASER to DM and the baryon asymmetry of the Universe.

%%%%%%%%%%%%%%%%%%%%%%%%%%%%%
\subsection{Left-right symmetric models\label{sec:LR}}

In the left-right (LR) symmetric models~\cite{Pati:1974yy, Mohapatra:1974gc, Mohapatra:1980yp}, the SM gauge group is extended to $SU(2)_L\times SU(2)_R\times U(1)_{B-L}$ with gauge couplings of left and right $SU(2)$ groups denoted by $g_L$ and $g_R$, respectively. As a consequence of the so-called left-right symmetry, additional right-handed doublets of quarks and leptons that appear in the model are constructed analogously to the left-handed doublets of $SU(2)_L$. In particular, this naturally implies the existence of three right-handed neutrinos that complete the right-handed lepton doublets and can lead to nonzero masses of the active neutrinos via the seesaw mechanism. The LR symmetry group is broken to the SM gauge group, $SU(2)_L\times U(1)_Y$, at some energy above the scale of EWSB. This leads to additional charged and neutral gauge bosons denoted by $W_R$ and $Z_R$, respectively. Current exclusion bounds imply $m_{W_R}\gg m_{W_L}$, while $Z_R$ is even heavier and we will henceforth neglect its impact.

Right-handed neutrinos in this model can be produced and subsequently decay in various processes involving $W_R$ exchange. Assuming, for simplicity, that $g_L=g_R$ and given that the mixing matrix in the right-handed quark sector is expected to follow closely the left-handed one~\cite{Senjanovic:2015yea}, the matrix elements of the relevant processes are rescaled with respect to the SM neutrinos by $(m_{W_L}/m_{W_R})^2$. Hence, phenomenology of the right-handed neutrinos in LR symmetric models closely resembles the above discussion of HNLs with a simple substitution $|U_{\ell N}|^2\rightarrow (m_{W_L}/m_{W_R})^4$. This allows us to easily translate limits on HNLs from various intensity frontier searches into the limits in the $(m_{W_R},m_N)$ plane (see, \eg,~\cite{Castillo-Felisola:2015bha}). 

Importantly, sterile neutrinos in the LR symmetric models with sizable mixing with the electron neutrino are strongly constrained by null searches of neutrinoless double-beta decay~\cite{Hirsch:1996qw,Deppisch:2012nb}. For this reason, we present the results for LR sterile neutrinos from the muon doublet assuming that other mixing angles are suppressed. In \Figref{LR}, we present the sensitivity reach of FASER in the $(m_{W_R},m_N)$ plane for such a scenario. As can be seen, FASER could provide limits reaching $m_{W_R}\gtrsim 7~\tev$ for a low mass of $N$ that would be complementary to the region of the parameter space probed in search for $W_R \to \mu N \to \mu\mu jj$ events in the high-$p_T$ searches at the LHC. In particular, we show the current limits for this channel obtained by the CMS Collaboration for $\sqrt{s}=13~\tev$ and $35.9~\ifb$ integrated luminosity~\cite{CMS:2017ilm}, as well as by the ATLAS Collaboration for $\sqrt{s}=8~\tev$ and $20.3~\ifb$ integrated luminosity~\cite{Aad:2015xaa}. Note that the ATLAS search also considers a topology with one jet in the final state extending their reach for light and highly boosted $m_N$. Additional constraints come from the dijet search $W_R \to jj$~\cite{Khachatryan:2015sja,Helo:2015ffa}. Further complementarity~\cite{Nemevsek:2018bbt} can be achieved by searches for multitrack displaced vertices~\cite{Helo:2013esa,Cottin:2018kmq}, neutrino jets~\cite{Mitra:2016kov} and even searches for $Z_R \to \ell\ell$~\cite{Lindner:2016lpp}. We also show for comparison the projected sensitivity of the proposed SHiP detector obtained based on~\cite{Alekhin:2015byh}, which could extend the future bounds up to $m_{W_R}\sim 20~\tev$ for a narrow range of the HNL masses. Other bounds, which are not shown in \Figref{LR}, from searches for HNLs that we have discussed in \secref{reach} can also constrain the parameter space of the LR symmetric models for low values of $m_{W_R}$. 

\begin{figure}[t]
\centering
\includegraphics[width=0.5\textwidth]{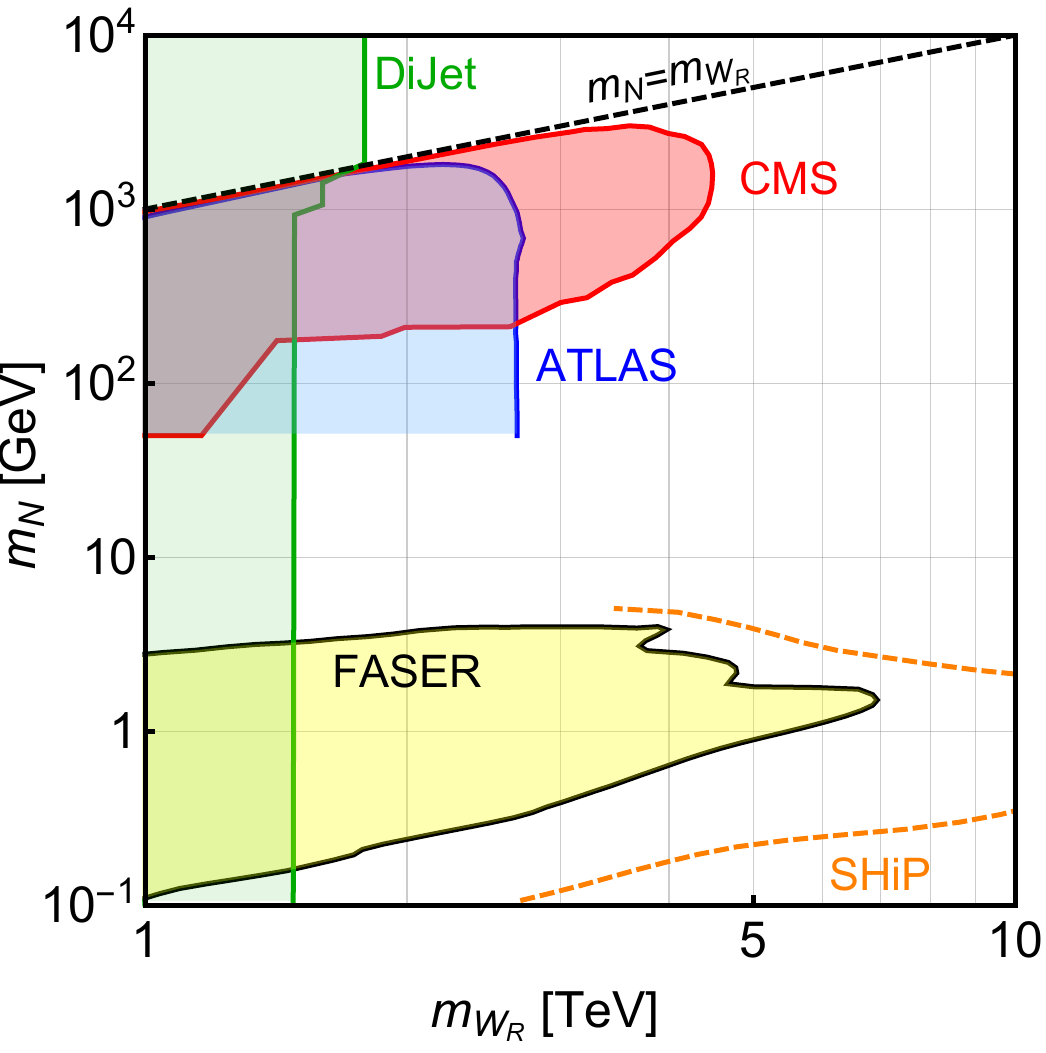}
\caption{Sensitivity reach of FASER to search for sterile neutrino in LR symmetric models in scenario with $U_{\mu N}\neq 0$ (yellow shaded area) in the $(m_{W_R},m_N)$ plane. Current limits~\cite{Aad:2015xaa,CMS:2017ilm} from CMS and ATLAS searches for sterile neutrinos and in $W_R \to \mu N \to \mu\mu jj $ decay are shown with red and blue shaded areas, respectively. The green shaded area corresponds to the limits in the dijet search $W_R \to jj$~\cite{Khachatryan:2015sja,Helo:2015ffa}. The projected reach of the SHiP experiment (dashed orange line) is also shown for comparison. In the region below the black dashed line, the heavy neutral lepton is lighter than $W_R$. For other bounds and future projections see the text.}
\label{fig:LR}
\end{figure}

%%%%%%%%%%%%%%%%%%%%%%%%%%%%%
\subsection{Sterile neutrinos from the mirror sector\label{sec:mirror}}

FASER can also be sensitive to scenarios employing other proposed mechanisms of generating the masses of the active neutrinos that go beyond type-I seesaw. Such a possibility with additional connections to DM and leptogenesis has been discussed in~\cite{An:2009vq} for the SM extended by its mirror sector. Notably, models with the mirror sector of the SM have recently received much attention also in the context of studies of \textsl{neutral naturalness} as a solution to the hierarchy problem. In particular, a prototypical such scenario is the twin Higgs model~\cite{Chacko:2005pe} with the SM duplicated in the mirror SM-singlet sector. It is needless to mention that neutral naturalness becomes increasingly motivated given lack of discovery of any colored BSM particles at the LHC.

In the model considered in~\cite{An:2009vq}, it is further assumed that both the SM and the mirror sector are connected via set of heavy right-handed Majorana neutrinos, $N_I$, that have Yukawa couplings to the SM leptons, as well as to their mirror counterparts
\begin{equation}
\mathcal{L} \supset -F_{\alpha I}\bar{L}_\alpha\,\widetilde{N}_I\,\tilde{\Phi} -F'_{\alpha I}\bar{L}'_\alpha\,\widetilde{N}_I\,\tilde{\Phi}'-\frac{1}{2}\bar{\widetilde{N}}^c_I\,M_{I}\,\widetilde{N}_I,
\label{eq:Lmirror}
\end{equation}
where the prime corresponds to the mirror sector. We will denote the active neutrino states in the SM and the mirror sector by $\nu$ and $\nu'$, respectively. In addition, for consistency, the set of Higgs bosons of the model has been extended to include Higgs doublets, $H_{u,d}$ ($H'_{u,d}$), and triplets, $\Delta$ ($\Delta'$), for the SM (mirror) sector. The latter couple to lepton pairs via $Y_\Delta\,\bar{L}^c\,\Delta\,L$ and $Y'_\Delta\,\bar{L'}^c\,\Delta\,L'$. The neutral components of both triplets can get a nonzero vacuum expectation value, which gives raise to Majorana (type-II seesaw) contributions to the neutrino mass matrix for $\nu$ and $\nu'$ that are denoted by $\mu$ and $\mu'$, respectively.

Leptogenesis in both sectors can be driven by out-of-equilibrium decays of the lightest of the right-handed neutrinos or by resonant processes if two right-handed neutrinos are mass degenerate. It is then transferred to the baryon asymmetry by sphaleron transitions. In particular, baryons of the mirror sector can play the role of DM. The correct ratio between baryon and DM abundances is set by the ratio between the SM and mirror proton masses. The latter is driven to the value $\sim 5\,m_p$ by assuming large scale of the EWSB in the mirror sector, $v_{\rm mir}/v_{\rm SM}\sim 10^3-10^4$, that makes mirror sector $u$ and $d$ quarks heavy.

After diagonalizing the neutrino mass matrix, the SM active neutrino masses have both the type-II seesaw contribution $\mu$ and the inverse seesaw contribution that can be written as $\mu'\,(v_{\rm SM}/v_{\rm mir})^2$.\footnote{The SM active neutrino masses can also arise in the absence of the Higgs triplet out of the radiative corrections as discussed in~\cite{Dev:2012sg,Zhang:2013ama}.} On the other hand, $\nu'$ masses in the mirror sector are mostly driven by type-I seesaw mechanism. Typically, two of them are in the $\sim\gev$ scale for the parameters of the model that allow successful leptogenesis and correct DM abundance~\cite{An:2009vq,Zhang:2013ama}. In addition, in the mass eigenstates, nonzero mixing between $\nu$ and $\nu'$ appears of order $v_{\rm SM}/v_{\rm mir}\sim 10^{-4}-10^{-3}$. As a result, $\nu'$s play the role of the light sterile neutrinos with properties resembling those of HNLs discussed above. Importantly, their typical masses and mixing angles lie in the region of the parameter space covered by FASER, which should, therefore, be able to test this interesting cosmological scenario. 

%%%%%%%%%%%%%%%%%%%%%%%%%%%%%
%%% Conlusion
%%%%%%%%%%%%%%%%%%%%%%%%%%%%%

\section{Conclusions\label{sec:conclusions}}

Searches for light long-lived particles have recently become widely accepted as one of the most important aims to achieve in the next generation of dedicated experiments looking for BSM physics. Among many models of new physics that can be probed in such searches, one of the most extensively studied scenarios predicts heavy neutral leptons with masses of order $\gev$ and small mixing angles with the active neutrinos. In particular, HNLs appear naturally in the context of the seesaw mechanism that generates nonzero neutrino masses in the SM. Although typically the seesaw scenario predicts sterile neutrinos with large masses of order the GUT scale, it is also possible to obtain light HNLs in theoretically appealing models.

In this study, we have analyzed the expected sensitivity reach to HNLs of the newly proposed detector to be placed along the LHC beam axis, named ForwArd Search ExpeRiment, or FASER~\cite{Feng:2017uoz,Feng:2017vli}. This analysis is complementary to our previous studies for dark photons that are mainly produced in pion decays and for dark Higgs bosons that typically come from $B$-meson decays. In the case of HNLs also decays of $D$ mesons play an important role, while $B$ mesons start to dominate only when other channels are kinematically forbidden. 

We compare the sensitivity reach of FASER in the HNL parameter space to other proposed detectors. In particular, we show that for sterile neutrinos produced mainly in $B$-meson decays, even a relatively small cylindrical detector with $\sim 10~\m$ length and $R\lesssim 1~\m$ radius can have reach comparable to the proposed SHiP detector, while other experiments typically lose their sensitivity in this range of the HNL mass. In this mass regime the current bounds in the mixing angles could be improved by about an order of magnitude. On the other hand, the reach in the mixing angle for HNLs produced in $D$-meson decays is comparable to the one expected from the NA62 experiment and can be surpassed by SHiP and DUNE due to their much larger luminosities. 

However, even in the case of $D$-meson decays, FASER can probe interesting region in the HNL parameter space. In particular, as discussed in \secref{mirror}, it corresponds to the attractive cosmological scenario in which sterile neutrinos emerge from the mirror sector of the SM with heavy right-handed neutrinos as mediators. In this scenario, the search for sterile neutrinos in FASER could shed light on other important questions in contemporary physics including the nature of DM and the origin of baryon asymmetry of the Universe. It could also be related to the increasingly popular solution to the hierarchy problem via neutral naturalness.

Importantly, looking for a signal from light sterile neutrino decays in FASER will play a role complementary to searches for heavier HNLs in high-$p_T$ oriented experiments at the LHC. We illustrate this in \secref{LR} for the case of the left-right symmetric models that provide a natural extension of the SM gauge group to explain the asymmetry between the left and right fermions in the SM.

Last, but not least, in scenarios with sterile neutrinos emerging from the mirror sector of the SM, the photon of the mirror sector mixes with the SM photon~\cite{An:2009vq,Zhang:2013ama}. It can then effectively play the role of the dark photon. Its mass and mixing parameter would then typically be within the reach of FASER discussed in our previous study~\cite{Feng:2017uoz}. In this case, one expects to simultaneously see events with the HNL and dark photon origin with possibly distinct final states of their decays.

%%%%%%%%%%%%%%%%%%%%%%%%%%%%%
% ACKNOWLEDGMENTS
%%%%%%%%%%%%%%%%%%%%%%%%%%%%%
\acknowledgments

We thank Jonathan Feng for many helpful remarks and for carefully reading the manuscript. We thank Iftah Galon for productive discussions at an early stage of the project. We thank Brian Shuve and Zeren Simon Wang for useful discussions. We thank Mike Lamont for invaluable support in analyzing the LHC infrastructure and in determining possible locations for FASER. This work is supported in part by NSF Grant No.~PHY-1620638. S.T.\ is supported in part by the Polish Ministry of Science and Higher Education under research grant 1309/MOB/IV/2015/0 and by the National Science Centre (NCN) in Poland research grant No.\ 2015-18-A-ST2-00748.  

%\newpage
\appendix
%%%%%%%%%%%%%%%%%%%%%%%%%%%%%
%%% Appendix Fragmentation
%%%%%%%%%%%%%%%%%%%%%%%%%%%%%
\section{Fragmentation into $D$ and $B$ mesons\label{sec:fragmentation}}

As discussed in Sec. III, the dominant contribution to production of HNLs in the context of FASER comes from decays of pseudoscalar D and B mesons. Below we discuss how the corresponding quark fragmentation is performed.
%%%%%%%%%%%%%%%%%%%%%%%%%%%%%
\subsection{$D$-mesons Fragmentation}

\begin{figure}[h]
\centering
\includegraphics[width=0.80\textwidth]{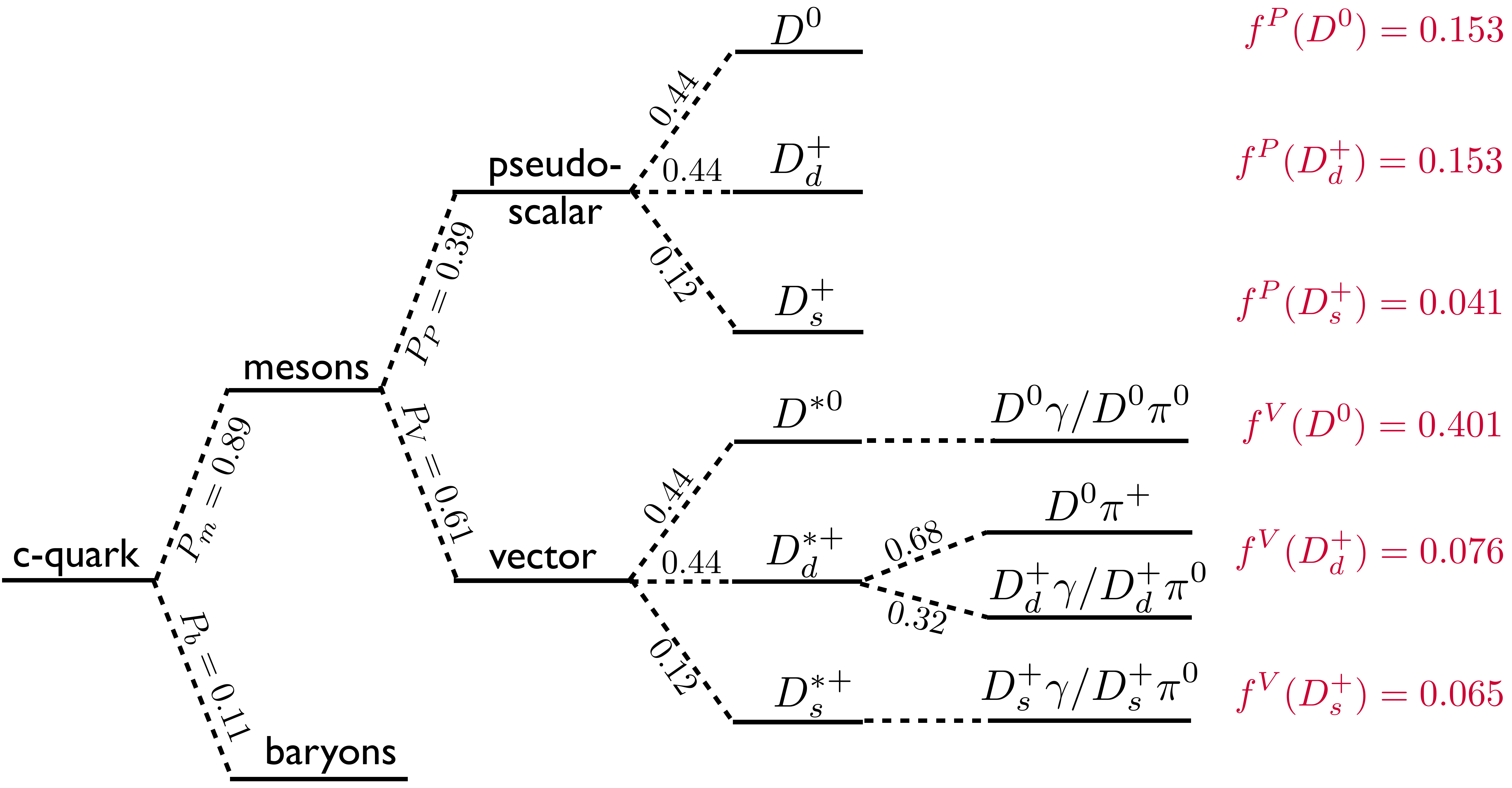}
\caption{Fragmentation tree for $c$ quarks. In red, we show the $c$-quark pseudoscalar and vector fragmentation fractions for $D^0$, $ D^+$, and $D_s^+$ mesons.}
\label{fig:Dfragtree}
\end{figure}

%As discussed in \secref{prodanddec}, the dominant contribution to production of HNLs in the context of FASER comes from decays of pseudoscalar $D$ and $B$ mesons. 
To determine the $D$-meson fragmentation fractions and functions we follow~\cite{PDGDfrag}. The fragmentation tree for charm quarks is shown in \Figref{Dfragtree}. 

The charm-quark fragments into mesons with a probability of $P_m = 0.89$ or baryons with $P_b = 0.11$. The baryonic mode is dominated by the production of $\Lambda_c$ with a $~14\%$ contribution from charmed-strange mesons $\Xi_c$ and $\Omega_c$~\cite{Chekanov:2005mm}. The mesons are either produced in the pseudoscalar or vector state with production fractions $P_P=0.39$ and $P_V = 0.61$, respectively~\cite{PDGDfrag}. The fragmentation fraction into the strange meson states $D_s^{(*)+}$ is given by $p_s=\gamma_s/(2+\gamma_s) \approx 0.12$, where we used strangeness suppression factor $\gamma_s=0.259$~\cite{PDGDfrag}. For $u$ and $d$ quarks, the fragmentation fractions are obtained from $p_{u,d} \approx (1-p_s)/2 \approx 0.44$, provided that the ratio of neutral and charged mesons is equal to $R_{u/d} = 1.02\approx 1$~\cite{PDGDfrag}. The vector-meson states will then decay into the pseudoscalar state and either a photon or pion. The dominant decay modes of the $D^{*+}_d$ meson into $D^0$ or $D_d^+$ have branching fractions $B(D^{*+}_d\to D^0 \pi^+)=0.68$ and $B(D^{*+}_d\to D^+_d (\pi^0/\gamma))=0.32$, respectively~\cite{Patrignani:2016xqp}. These numbers are then combined into pseudoscalar and vector fragmentation fractions, $f^P$ and $f^V$, for the pseudoscalar mesons $D^0$, $D^+_d$, and $D^+_s$ as shown on the right side of \Figref{Dfragtree}. 

The pseudoscalar and vector fragmentation fractions $f^{P,V}$ for each meson have different nonperturbative fragmentation functions $F^P(z)$ and $F^V(z)$, respectively, which depend on the momentum fraction $z$. We employ the BCFY fragmentation functions~\cite{Braaten:1994bz} with default FONLL settings~\cite{Cacciari:2012ny}: $m_c=1.5~\gev$, $N=5$ and $r=0.1$. The total fragmentation function for each meson is given by
\be
F (D_i) = f^{V}(D_i) F^{V}(D_i) + f^{P}(D_i) F^{P}(D_i),
\ee
where $D_i=D^0$ and $D^+_d$ and $D^+_s$ are the pseudoscalar $D$ mesons for which we obtain the kinematic distributions.

%%%%%%%%%%%%%%%%%%%%%%%%%%%%%
\subsection{$B$-mesons Fragmentation}

To simulate $B$-meson fragmentation, we use a nonperturbative fragmentation function which follows the distribution by Kartvelishvili et al.~\cite{Kartvelishvili:1977pi,Cacciari:2005uk} with fragmentation parameter $\alpha = 24.2$. At this stage, we do not differentiate between various $B$-meson final states. 

To determine the number of various $B$ hadrons, we employ fragmentation fractions $f_u, f_d, f_s, f_c$ and $f_\Lambda$ that correspond to $B^+_u, B^0_d, B^0_s, B^+_c$, and $\Lambda_b$ hadrons in the final state. For simplicity, we include all baryonic states in $f_\Lambda$ so that the above fragmentation fractions sum up to unity. We then employ the LHCb results for $f_i$~\cite{Aaij:2011jp,Aaij:2017kea} that correspond to the continuum $b\bar{b}$ production mode 
\be
\frac{f_d}{f_u}\approx 1,\quad \frac{f_s}{f_u+f_d}=0.135 \pm 0.01 ,\quad \frac{f_c}{f_u} =0.008 \pm 0.004, \quad \frac{f_{\Lambda}}{f_u + f_d}=0.404 \pm 0.11.
\ee
Note that we only take into account results from LHCb, which use the continuum $\bar{b}b$ production mode. These results agree very well with the fragmentation spectrum measured at the Tevatron. The fragmentation fractions measured at BaBar and Belle at the $\Upsilon$ resonance and LEP at the $Z$ resonance differ slightly due to the resonant production mode. We can solve for the individual fragmentation fractions $f_i$ and obtain
\be
f_u=0.324,\quad f_d=0.324,\quad  f_s=0.0875,\quad f_c=0.0026,\quad f_\Lambda=0.262.
\ee

\bibliography{HNL}

\end{document}